%% file: main.tex
\documentclass[prl,reprint,preprintnumbers,longbibliography]{revtex4-2}
\usepackage{textgreek}

\input{preamble.tex}
\input{acronyms.tex}
\renewcommand\O[1]{\text{O}(#1)}

\begin{document}
\title{
  \texorpdfstring{%
    From asymptotic freedom to $\theta$ vacua:\\ Qubit embeddings of the O(3) nonlinear $\sigma$ model}%
  {From asymptotic freedom to θ vacua: Qubit embeddings of the O(3) nonlinear sigma model}
}

\author{Stephan Caspar\,\orcidlink{0000-0002-3658-9158}}
\email{caspar@uw.edu}
\author{Hersh Singh\,\orcidlink{0000-0002-2002-6959}}
\preprint{IQuS@UW-21-025, INT-PUB-22-012}
\email{hershsg@uw.edu}
\affiliation{InQubator for Quantum Simulation (IQuS), Department of Physics, University of Washington, Seattle, Washington 98195-1550, USA}

\begin{abstract}
  Conventional lattice formulations of $\theta$ vacua in the $1+1$-dimensional $\O3$ nonlinear sigma model suffer from a sign problem. Here, we construct the first sign-problem-free regularization for \emph{arbitrary} $\theta$. Using efficient lattice Monte Carlo algorithms, we demonstrate how a Hamiltonian model of spin-$\tfrac12$ degrees of freedom on a 2-dimensional spatial lattice reproduces both the infrared sector for arbitrary $\theta$, as well as the ultraviolet physics of asymptotic freedom. Furthermore, as a model of qubits on a two-dimensional square lattice with only nearest-neighbor interactions, it is naturally suited for studying the physics of $\theta$ vacua and asymptotic freedom on near-term quantum devices. Our construction generalizes to $\theta$ vacua in all $\CP(N-1)$ models, solving a long standing sign problem.
\end{abstract}

\maketitle
\glsresetall

\input{sec_intro}
\input{sec_theory}
\input{sec_results}
\input{sec_conclusions}

\section*{Acknowledgments}

We learned about D-theory from Shailesh Chandrasekharan and Uwe-Jens Wiese and are grateful to them for many enlightening conversations over the years.
We thank Martin Savage for inspiring discussions and important feedback on the manuscript.
We would also like to acknowledge stimulating conversations with
Tanmoy Bhattacharya, Anthony Ciavarella, Mendel Nguyen and Mithat Ünsal on related matters.

The material presented here was funded 
in part by the DOE QuantISED program through the theory  consortium ``Intersections of QIS and Theoretical Particle Physics'' at Fermilab with Fermilab Subcontract No. 666484,
in part by Institute for Nuclear Theory with US Department of Energy Grant DE-FG02-00ER41132,
and in part by U.S. Department of Energy, Office of Science, Office of Nuclear Physics, Inqubator for Quantum Simulation (IQuS) under Award Number DOE (NP) Award DE-SC0020970.

\bibliography{refs}

\end{document}

%% file: preamble.tex
\usepackage{multirow}
\usepackage{siunitx}

\usepackage{tikz}
\usetikzlibrary{decorations.markings, arrows, positioning, calc}

\usepackage{graphicx,times}
\usepackage{latexsym}
\usepackage{mathtools}

\usepackage{amsmath,amssymb,amsbsy,amsfonts}
\usepackage{array}
\usepackage{bm}
\usepackage{graphics}
\usepackage{mathrsfs}
\usepackage{xcolor}
\usepackage{cancel}
\usepackage[normalem]{ulem}

\usepackage[unicode, pdfprintscaling=None, colorlinks]{hyperref}
\hypersetup{
    colorlinks,
    allcolors=blue!50!black,
}
\usepackage{inconsolata}

\usepackage[capitalise]{cleveref}

\newcommand{\RR}{{\mathbb{R}}}

\newcommand\Order{O}

\DeclareMathOperator\Tr{Tr}

\usepackage{relsize}

\usepackage{microtype}

\usepackage{xfrac}
\usepackage{orcidlink}
\newcommand\SU{\ensuremath{\text{SU}}}
\newcommand\CP{\ensuremath{\text{CP}}}

\usepackage{ulem} 
\usepackage{amssymb}

\newcommand\Lx{\ensuremath{L_X}}
\newcommand\Ly{\ensuremath{L_Y}}
\newcommand\Lt{\ensuremath{L_T}}

\newcommand\del\partial

%% file: acronyms.tex
\usepackage{relsize}

\usepackage[acronyms, smallcaps, nohypertypes={acronym}, nopostdot, style=super, nonumberlist, toc]{glossaries}
\glsenableentrycount
\makeglossaries  
\setacronymstyle{long-sc-short}

\newcommand\ac[1]{\gls{#1}}

\newcommand\acp[1]{\glspl{#1}}

\newacronym{WF}{wf}{Wilson-Fisher}
\newacronym{AF}{af}{asymptotically free}
\newacronym{RG}{rg}{renormalization group}
\newacronym{WZW}{wzw}{Wess-Zumino-Witten}
\newacronym[longplural={conformal field theories}]{CFT}{cft}{conformal field theory}
\newacronym[longplural={lattice field theories}]{LFT}{lft}{lattice field theory}
\newacronym[longplural={effective field theories}]{EFT}{eft}{effective field theory}
\newacronym[longplural={quantum field theories}]{QFT}{qft}{quantum field theory}
\newacronym{LEC}{lec}{low-energy constant}
\newacronym{QCD}{qcd}{quantum chromodynamics}
\newacronym{MC}{mc}{Monte Carlo}
\newacronym{IR}{ir}{infrared}
\newacronym{UV}{uv}{ultraviolet}
\newacronym{SNR}{snr}{signal-to-noise ratio}
\newacronym{NLSM}{nl$\sigma$m}{nonlinear sigma model}
\newacronym{CSA}{csa}{Cartan subalgebra}
\newacronym{SSB}{ssb}{spontaneous symmetry breaking}
\newacronym{DOF}{dof}{degrees of freedom}
\newacronym{DMRG}{dmrg}{densiy matrix renormalization group}

%% file: sec_intro.tex
\section{Introduction}

The strong interactions of the standard model described by \ac{QCD} pose a challenging problem for classical computation. While nonperturbative lattice \ac{MC} methods are a powerful tool for studying static properties of strongly coupled \acp{QFT} like \ac{QCD}
\cite{ratti_lattice_2018, davoudi_nuclear_2021,hansen_lattice_2019, constantinou_parton_2021}, questions involving real time dynamics, finite density or nontrivial $\theta$ vacua are still out of reach for lattice \ac{MC} methods due to  severe sign problems \cite{hsu2010sign,goy2017sign}.

Emerging quantum platforms provide an exciting possibility for investigating \acp{QFT} in previously inaccessible regimes. 
They are not directly affected by the sign problems arising in classical lattice \ac{MC} methods. 
However, bosonic lattice field theories such as \ac{QCD} have infinite-dimensional local Hilbert spaces, while  hardware \ac{DOF} are 
usually finite-dimensional, mostly qubits. 
A significant effort is underway to explore different embeddings of \acp{QFT} in qubits, with a multitude of ideas emerging from bosonic field theory \cite{jordan2011quantum,jordan2012quantum,yeter2019scalar,klco2019digitization}, \acp{NLSM} \cite{chandrasekharan_spin_2002,brower_dtheory_2004, beard_efficient_2006,laflamme_cp_2016, evans_su_2018a,bruckmann_nonlinear_2019,singh_qubit_2019,bhattacharya_qubit_2021, singh_qubit_2019a, singh_largecharge_2022a} and gauge theories \cite{chandrasekharan_quantum_1997a, brower_qcd_1999,raychowdhury2020solving,anishetty2009prepotential,banerjee2013atomic,zohar2015quantum,banuls2017efficient,muschik2017u,zache2018quantum,alexandru2019gluon,bender2020gauge,davoudi2020towards,klco20202,shaw2020quantum,kasper2020non,buser2021quantum,haase2021resource}.

The $1+1$-dimensional $\O3$ \ac{NLSM} has a long history  as a prototype for \ac{QCD}, due to similarities such as asymptotic freedom, dynamical transmutation and the generation of a nonperturbative mass gap, as well as a topological $\theta$ term. 
The $\O3$ \ac{NLSM} with a $\theta$-term is formally defined by the continuum action
\begin{align}
  S_\theta[\vec \phi] = \frac{1}{g^2} \int d^2  x (\partial_\mu \vec \phi)^2 + i \theta Q[\vec \phi],
  \label{eq:Stheta}
\end{align}
where $\vec{\phi} \in \RR^3$ with $|\vec \phi|^2=1$, and
\begin{align}
  Q[\vec \phi] = \frac{1}{8\pi} \int d^2 x\ \varepsilon_{\mu \nu}\  \vec \phi \cdot (\del^\mu \vec \phi) \times  (\del^\nu \vec \phi)
  \label{eq:TopC}
\end{align}
is the integer topological charge, making the theory $2 \pi$-periodic in $\theta$.
Both $\theta=0,\pi$  points are well-understood, analytically as well as on the lattice.
Exact S-matrices have been conjectured for both $\theta=0$ and $\theta=\pi$ \cite{zamolodchikov_factorized_1979, balog_tba_2004, balog_finite_2010, luscher_quantum_1978} and their integrability has been confirmed using non-perturbative lattice \ac{MC} methods \cite{luscher_numerical_1991,bietenholz_testing_1996,caracciolo_asymptotic_1995}.

\begin{figure}
    \centering
    \includegraphics{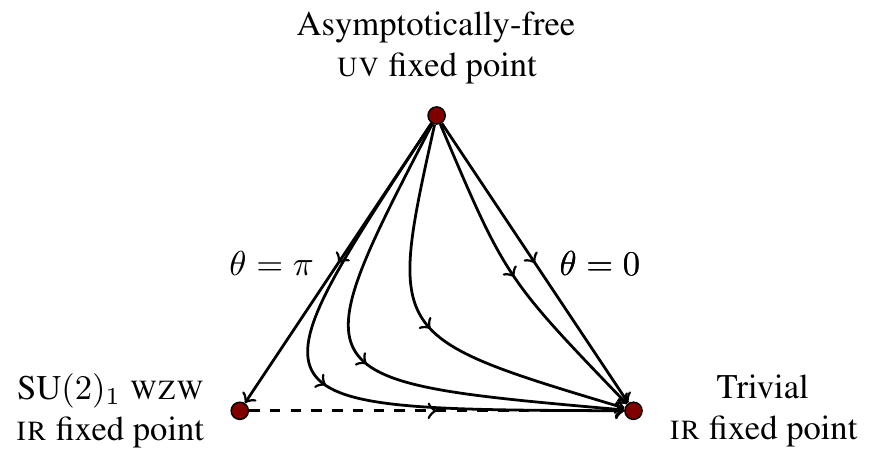}
    \caption{\textsc{rg} flow diagram of $\O3$ \acp{NLSM} $S_\theta$ defined in \cref{eq:Stheta}. $S_\theta$ is a family of asymptotically-free \acp{QFT} which all flow into the trivial \ac{IR} fixed point, except at $\theta = \pi$ where it reaches the $\SU(2)_1$ $\ac{WZW}$ fixed point. At small $|\theta-\pi|$, the \ac{RG} flow of $S_\theta$ passes arbitrarily close to the \ac{WZW} fixed point, on its way to the trivial fixed point. }
    \label{fig:rg}
\end{figure}

However, general, non-integrable $\theta$ remain challenging.
As a topological effect, it cannot be studied directly in perturbation theory about the free \ac{UV} fixed point, although 
some analytic progress has been made by perturbing about the $\theta=\pi$ integrable point \cite{balog_twodimensional_2000}. 
Nonperturbatively, the inclusion of a $\theta$ term causes a sign problem when discretizing the action in \cref{eq:Stheta} on a $2$-dimensional spacetime lattice. Even though improved actions combined with cluster algorithms have been shown to tame both cutoff effects and the sign problem to allow a reliable extrapolation from modest volumes around $\theta \approx 0$ \cite{bogli_nontrivial_2012, deforcrand_walking_2012}, and even fully solve the sign problem at $\theta = \pi$ \cite{bietenholz_testing_1996}, 
so far there are no known lattice \ac{MC} methods which allow a fully controlled study of arbitrary $\theta$ vacua.

Motivated by the prospect of quantum simulation to address these challenges, we develop an embedding of the $\O3$ \ac{NLSM} at \emph{arbitrary} $\theta$ into a $2$-dimensional Heisenberg antiferromagnet, such that a controlled continuum limit can be taken.
Remarkably, not only does this model allow the systematic study of $\theta$ vacua on quantum hardware, it also enables the first sign problem free algorithm for classical computations at arbitrary $\theta$. This extends a similar proposals put forward in Refs.~\cite{chandrasekharan_spin_2002, chandrasekharan_quantum_1997a, brower_dtheory_2004,laflamme_cp_2016} for classical and quantum simulation of $\theta = 0,\pi$ theories.

Regularizing \acp{QFT} using explicitly finite-dimensional local \ac{DOF} is a promising approach for quantum simulation. Universality lets us understand this remarkable variety in models with the ability to describe the same continuum \acp{QFT}. In his seminal work on \ac{RG}, Wilson showed how continuum \acp{QFT} emerge at second-order critical points of lattice models \cite{wilson_renormalization_1974, wilson_renormalization_1983, kogut_introduction_1979}. 
In this framework, the infinite-dimensional continuum fields can arise naturally at long-distances from finite-dimensional microscopic local \ac{DOF}. While this approach is natural in the context of quantum computation, universality has even been leveraged to circumvent sign problems that plague conventional lattice regularizations. 
This was shown, for example, with the $\O3$ model at finite density \cite{chandrasekharan_spin_2002} and the $\CP(2)$ model at $\theta=\pi$ \cite{beard_study_2005}.
Efficient cluster algorithms for $\CP(N-1)$ models have been demonstrated
\cite{beard_efficient_2006}, where a no-go theorem prevents efficient cluster algorithms using the standard lattice action \cite{caracciolo_wolff_1993}.

%%% Local Variables:
%%% mode: latex
%%% TeX-master: "main"
%%% End:

%% file: sec_theory.tex
\begin{figure}
  \centering
  \includegraphics[width=0.95\linewidth]{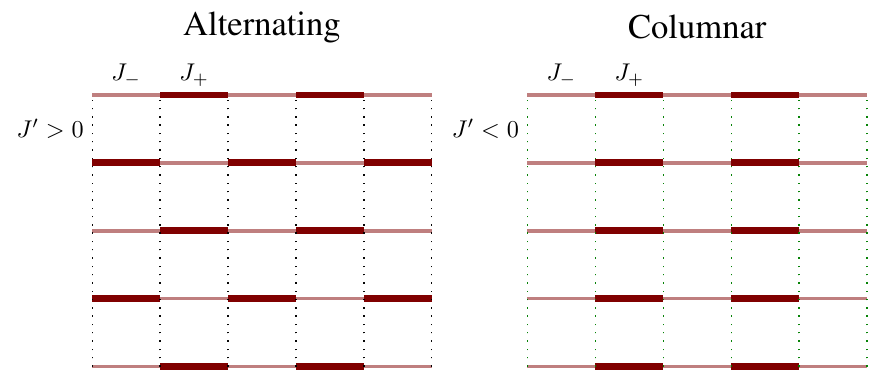}
  \caption{%
    Two configurations for the staggered interactions, described in \cref{eq:staggering}, considered as a regularization of the $1+1$-dimensional $\O3$ \ac{NLSM} with a $\theta$ term.
    For the alternating staggering, all couplings are antiferromagnetic, while for the columnar case, the transverse coupling $J'$ is ferromagnetic and $J_\pm = J(1 \pm \gamma)$ is antiferromagnetic. All interactions are of the Heisenberg $\vec S_i \cdot \vec S_j$ type.
    }
  \label{fig:staggering}
\end{figure}

\section{The Qubit Hamiltonian}

In this work, we show that the continuum limit of the $1+1$d $\O3$ \ac{NLSM} with a $\theta$ term can be obtained from a spin-$\frac12$ Heisenberg antiferromagnet on a $2d$ lattice with staggered couplings
\begin{align}
  H =
  \sum_{(x,y)} J_{x,y}\ \vec S_{x,y} \cdot \vec S_{x+1,y}
  + J' \sum_{ (x,y) } \ \vec S_{x,y} \cdot \vec S_{x,y+1}.
  \label{eq:DTHam}
\end{align}
where $\vec S_i$ are the spin operators acting on two-dimensional Hilbert space at the site $(x,y)$,
$J_{x,y}$ are the couplings along the $x$ direction, $J'$ is the coupling along the $y$ direction, and the $2d$ lattice has dimensions $\Lx \times \Ly$.
We consider the following two configurations for staggering the couplings:
\begin{align}
  \text{Alternating:} && J' > 0, \quad J_{x,y} &= J(1 + (-1)^{x+y} \gamma),
  \label{eq:staggering} \\
  \text{Columnar:} && J' < 0, \quad J_{x,y} &= J(1 + (-1)^{x} \gamma),
  \nonumber
\end{align}
where $J>0$ is always antiferromagnetic, and $\gamma$ is the staggering parameter, as shown in \cref{fig:staggering}. 
In both these cases, the continuum limit of the $\O3$ \ac{NLSM} with a $\theta$ term can be obtained
from odd or even $\Ly$, by taking the limit $\Ly \to \infty$ at fixed $\gamma\Ly$ such that $\Lx \gg \Ly \gg 1$ is maintained.

To demonstrate the continuum limit, we need to recover the physics of the theory described by \cref{eq:Stheta} at all scales, from the \ac{UV} to the \ac{IR}.
For all $\theta$, the continuum action $S_\theta$, defined in \cref{eq:Stheta}, describes an asymptotically-free theory, controlled in the \ac{UV} by the fixed point of two free bosons.
The coupling $g$ is a relevant coupling and thus drives the theory away from the free \ac{UV} fixed point into a strongly coupled theory in the \ac{IR}.
While all $S_\theta$ theories flow out of the same \ac{UV} fixed point, non-perturbative effects lead to different \ac{RG} trajectories for different $\theta$.
\Cref{fig:rg} shows a conjectured \ac{RG} flow diagram for the $\O3$ \ac{NLSM} at arbitrary $0 \leq \theta \leq \pi$.
For all $\theta \neq \pi$ the theory flows to the trivial massive fixed point in the \ac{IR}.
However, at $\theta=\pi$, the theory undergoes a second order phase transition and the low-energy physics changes completely.
The mass-gap vanishes and the \ac{IR} physics is described by a nontrivial \ac{CFT} called the $\SU(2)_{1}$ \ac{WZW} theory \cite{shankar_th_1990}. 
Interestingly, the two ideas of staggering \cite{affleck_field_1988, martin-delgado_phase_1996} and D-theory \cite{brower_dtheory_2004} can be combined with the qubit Hamiltonian of \cref{eq:DTHam} to reproduce the physics of both \ac{IR}  and \ac{UV}.

\begin{figure}
  \centering
  \includegraphics{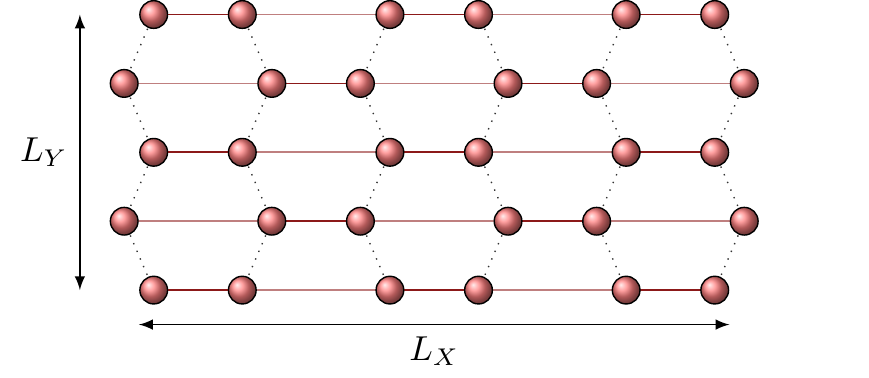}
  \caption{%
    Proposed embedding of the $\O3$ \ac{NLSM} with a $\theta$ term into a 2-dimensional array of ultracold atoms.
    The alternating staggering described in \cref{eq:staggering} and \cref{fig:staggering} arises naturally from distance-dependent antiferromagnetic interactions by deforming a rectangular lattice.
  }
  \label{fig:analog}
\end{figure}

\emph{IR physics of the $\theta$ vacua}.
For $\gamma=0$, the Hamiltonian  of \cref{eq:DTHam} reduces to the ordinary Heisenberg antiferromagnet.
For a fixed $\Ly$ and $\gamma=0$, this model has been studied in condensed matter literature as spin ladders, and is known to be described by the $\O3$ \ac{NLSM} at low-energies with $\theta = 2\pi S \Ly$
\cite{sierra_nonlinear_1996, haldane_continuum_1983, haldane_nonlinear_1983}.
Under this identification, the translation-by-one symmetry of spins ($\vec S_{x,y} \mapsto \vec S_{x+1,y}$) on the lattice scale becomes the charge conjugation symmetry ($\vec \phi \mapsto - \vec \phi$) in the continuum. Therefore, a $\theta$-term can be induced in the \ac{IR} by introducing a staggered coupling which breaks this symmetry \cite{martin-delgado_phase_1996, affleck_critical_1987, affleck_field_1988}.  Ref.~\cite{martin-delgado_phase_1996} showed that for spin-$S$ ladders with alternating staggering $\theta = 2\pi S \Ly (1 + \gamma f(\Ly))$, where $f(\Ly)$ is a non-universal function.
Therefore the low-energy physics of the $\theta$ vacua can be studied by varying the staggering parameter $\gamma$ \cite{venuti_particle_2005}.
However, to obtain the continuum limit of the $\O3$ \ac{NLSM}, we must also obtain the physics of asymptotic freedom in the \ac{UV}, which we now turn to.

\emph{Regulating the UV: Asymptotic Freedom.} 
The continuum limit of $S_\theta$, in \cref{eq:Stheta}, can be obtained from this Hamiltonian model by considering the limit $\Ly \to \infty$ while maintaining $\Lx \gg \Ly \gg 1$. This approach has been developed under the name D-theory \cite{chandrasekharan_spin_2002, brower_dtheory_2004, chandrasekharan_quantum_1997a}, and works as follows.
In the thermodynamic limit $\Lx, \Ly \to \infty$, the ground state of the Heisenberg antiferromagnet has Néel ordering with spontaneously broken global $SU(2)$-symmetry, with massless Goldstone mode excitations.
As $\Ly$ is made finite, the system develops an exponentially large correlation length $\xi \sim e^{\# \Ly} \gg \Ly$. 
The physics is therefore frozen along the $\Ly$ direction,
and the system effectively becomes one-dimensional, described by the $1+1$d $\O3$ \ac{NLSM} with an effective coupling $g^2 \sim 1/\Ly$. 
Since the correlation length diverges exponentially in $\Ly$, a continuum \ac{QFT} can be by defined in the limit of $\Ly$ large. 
Therefore, in this limit, the spin-$\frac12$ Hamiltonian of \cref{eq:DTHam} is  a lattice regularization of the $\O3$ \ac{NLSM} with an arbitrary $\theta$
at all scales, including asymptotic freedom in the \ac{UV}.

\emph{Extension to $\CP(N-1)$ models.} All methods in this paper are straightforward to extend from $\O3 = \CP(1)$ to the entire family of $\CP(N-1)$ models, which also allow for a $\theta$ term. 
Both $\theta=0,\pi$ have been considered before in the D-theory formulation \cite{laflamme_cp_2016, evans_su_2018a, beard_study_2005, beard_efficient_2006} 
using a Heisenberg model of $\SU(N)$ spins, where the $\SU(N)$ representations are chosen such that \ac{SSB} of the type $\SU(N) \to \text{U}(N-1)$ occurs \cite{read_valencebond_1989, read_features_1989}. 
This ensures that the continuum $\CP(N-1)$ fields arise as Goldstone modes as the continuum limit ($\Ly\to\infty$) is taken.
Since the discussion of charge conjugation symmetry is identical to that for the $\O3$ model, the staggering patterns of \cref{eq:staggering} will induce $\theta\neq 0,\pi$ in these constructions as well.

\begin{figure*}[ht!]
  \centering
  \includegraphics[trim=15 0 90 0,clip ,width=1.0\linewidth]{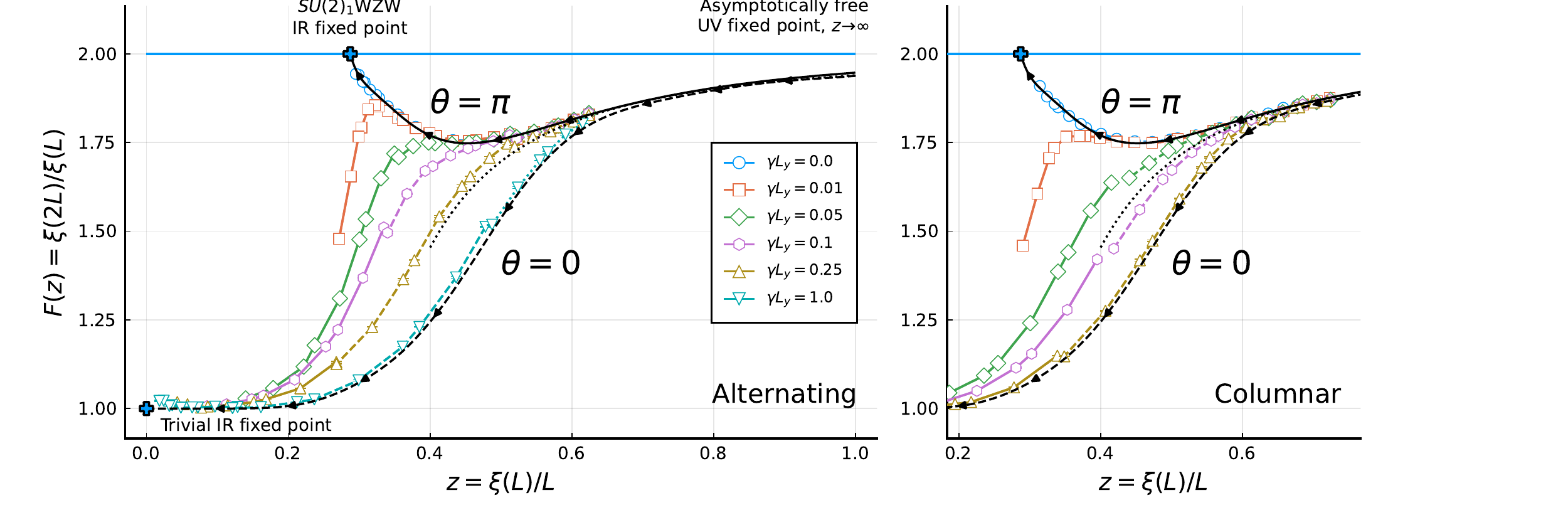}
  \caption{Step-scaling function of the $\O3$ \ac{NLSM} with various $\theta$.
    We show step-scaling curves for different values of $\gamma \Ly \sim |\theta-\pi|/\pi$ with odd $\Ly$, obtained from alternating (left) and columnar staggering (right), as defined in
    \cref{eq:staggering}.
    For a fixed $\gamma \Ly$, we show \ac{MC} results for $\Ly=3$ (solid line), $\Ly=5$ (dashed lines) and $\Ly=7$ (dotted lines).
    The dotted black curve is a two-loop perturbative prediction \cite{caracciolo_asymptotic_1995}.
    The dashed black line is the step-scaling function for $\theta=0$ obtained in Ref.~\cite{caracciolo_asymptotic_1995}.
    The solid black line is an $\Order(z^{-10})$ fit to the $\gamma \Ly = 0$ data, which corresponds to the step-scaling function of the $\O3$ \ac{NLSM} at $\theta=\pi$.
    These curves mimic the \ac{RG} flow diagram shown in \cref{fig:rg}, and 
    arrows on the $\theta=0,\pi$ curves indicate \ac{RG} flow from \ac{UV} to \ac{IR}.
    All curves agree in the perturbative \ac{UV} regime, while nonperturbative effects from the $\theta$ term lead to divergent trajectories in the \ac{IR}.
  }
  \label{fig:widessf}
\end{figure*}

%%% Local Variables:
%%% mode: latex
%%% TeX-master: "main"
%%% End:

%% file: sec_results.tex
\section{Methods}

In this work, we study the Hamiltonian defined in \cref{eq:DTHam} by performing \ac{MC} sampling of the partition function $Z = \Tr e^{-\beta H}$ using a worm algorithm \cite{prokofev_worm_2010, prokofev_worm_2001,wiese_determination_1994} on a spacetime lattice.
The dimensions $\Lx \times \Ly$ of the $2$-dimensional spatial lattices were varied in the range $32 \leq \Lx \leq 1024$ and $\Ly = 3,5,7$, with periodic boundary conditions in $\Lx$ and open boundary conditions in $\Ly$.
The couplings $J=|J'|=1$ were held fixed and only the staggering $\gamma$ was varied in the range $0\leq \gamma \Ly \leq 2$. The Heisenberg model with this level of staggering is not frustrated and no sign problem occurs in the \ac{MC} sampling. 
In our computations, the imaginary-time extent $\beta$ was also discretized into $\Lt$ timesteps of size $\varepsilon$, such that $\beta=\varepsilon \Lt $. Strictly speaking, the Hamiltonian model of \cref{eq:DTHam} is recovered only by extrapolating to the $\varepsilon \to 0$  limit. Alternatively, one can develop a cluster algorithm directly in the continuous time limit \cite{beard_simulations_1996}. However, since we are interested in studying the continuum limit of a relativistic field theory, we perform \ac{MC} computations at a fixed $\varepsilon=1.0$, which gives a transfer matrix model with the same continuum limit. The physical extent $\beta$ is chosen such that $\beta=\Lx/c$, where $c$ is the speed of light of this system.

\newcommand\gb{g_{\text{bare}}}
For all combinations of the parameters we calculate the second-moment correlation length $\xi_2(\Lx,g)$ from the spin-spin correlation function $\langle \vec{S}_{x,y} \cdot \vec{S}_{x',y'} \rangle$. This long distance length scale has been extensively studied at $\theta=0$ \cite{caracciolo_wolff_1993, caracciolo_asymptotic_1995} and is easy to extract from lattice results. The calculation is then repeated with doubled volume $2 \Lx$ but fixed bare couplings $\gb =(\Ly,J',J, \gamma)$. This macroscopic change in scale $\Lx \to 2 \Lx$ defines a discrete variant of the $\beta$-function, known as the step-scaling function \cite{luscher_numerical_1991}
\begin{align}
  F_\xi(z) &=\frac{\xi_2(2 \Lx, \gb)}{\xi_2(\Lx,\gb)}, &
  z=\xi_2(\Lx,\gb)/\Lx.
\end{align}
where $z$ defines a renormalized coupling.
In the continuum limit $\xi_2 \to \infty$, at constant $z$, the step-scaling function $F_\xi(z)$ becomes a universal function, which uniquely characterizes the corresponding \ac{QFT}.

\section{Results}

\Cref{fig:widessf} shows numerical results for the step-scaling function for the $\O3$ \ac{NLSM} at various $\theta$, computed using the qubit Hamiltonian of \cref{eq:DTHam}.
Results from both alternating (left panel) and columnar (right panel) staggering configurations are shown.
To guide the reader, we show three continuous curves: the perturbative prediction (dotted line) and nonperturbative \ac{MC} results for $\theta=0$ (black dashed line), and $\theta=\pi$ (black solid line).
The perturbative curve is a two-loop computation \cite{caracciolo_asymptotic_1995} valid in the \ac{UV} ($z \gg 1$) and shows asymptotic freedom near the \ac{UV} fixed point $F(z)=2$ at $z\to \infty$.
The $\theta = 0$ curve was obtained with the standard lattice action in Ref.~\cite{caracciolo_asymptotic_1995}, and shows the flow from the \ac{UV} to the trivial \ac{IR} fixed point ($F_\xi(0)=1$ at $z=0$).

The $\theta=\pi$ curve (black solid line) is a polynomial fit in $z^{-2n}$ up to order $n=5$ to our \ac{MC} results with $\gamma\Ly=0$.
This shows the \ac{RG} flow from the asymptotically-free \ac{UV} fixed point at $z=\infty$ to the $\SU(2)_1$ \ac{WZW} fixed point in the \ac{IR} at $z=z^*$.
We estimate the location of the nontrivial \ac{IR} fixed point to be $z^\ast \approx 0.28$ where $F(z^*) = 2$, which is the discrete equivalent of a vanishing $\beta$-function.
We emphasize that the physics  of \emph{all} scales, from asymptotic freedom in the \ac{UV} to the $\SU(2)_1$ \ac{WZW} theory in the \ac{IR} is reproduced by this model.

The remaining curves show new results for non-zero $\gamma \Ly \sim |\theta-\pi|/\pi$.
The $\theta=0,\pi$ curves form a lower and upper bound on all step scaling curves $0 \leq \theta \leq \pi$.
All curves closely follow the perturbative two-loop calculation (dotted line) at large $z$ down to $z\approx 0.75$.
At lower values of the renormalized coupling, non-perturbative effects start to dominate, leading to divergent trajectories.

For small staggering $\gamma \Ly$ the curves closely track the $\theta=\pi$ curve. 
But since $\theta$ is a relevant perturbation about the \ac{WZW} fixed point, the \ac{RG} trajectories cannot reach the nontrivial fixed point at $z^\ast$ and ultimately have to flow away to the trivial fixed point at $z=0$, consistent with the fact that these theories are massive.
These theories can be made to pass arbitrarily close to the $\SU(2)_1$ \ac{WZW} fixed point by choosing smaller and smaller $\gamma \Ly$, without the need for any fine tuning,
exemplified by the curve $\gamma \Ly = 0.01$ in \cref{fig:widessf}.
This is the phenomenon of conformal walking, which is also exhibited by \ac{QCD}-like 4-dimensional non-Abelian gauge theories near the conformal window \cite{banks_phase_1982}, or technicolor extensions of the Standard model \cite{deforcrand_walking_2012}.

As the staggering $\gamma$ is increased further, the step scaling curves trace out the entire area bounded by the two curves $\theta=0,\pi$, demonstrating that all $\theta$ vacua are contained in this model.
However, this only yields a qualitative relationship between $\gamma \Ly$ and $\theta$. Semi-classical results from large-$(S\Ly)$ expansions \cite{martin-delgado_phase_1996,sierra_application_1997} suggest that the relationship should be linear, $\theta = 2\pi S \Ly (1 + \gamma f(\Ly))$ with $f(\Ly) \to f(\infty)$ approaching a finite constant in the large-$\Ly$ limit.
Numerically, we observe that a value of $\gamma \Ly = 1.0$ ($\gamma \Ly = 0.25$) approximates the $\theta=0$ curve with the alternating (columnar) staggering.
Additional data also show a periodic reappearance of $\theta=\pi$ around values of $\gamma \Ly = 2.0$ ($\gamma \Ly = 0.5$).
From this we estimate the asymptotic values $f(\infty) \approx 1.0$ for alternating and $f(\infty) \approx 0.25$ for columnar staggering.
The small discontinuities of the step-scaling curves between different values of $\Ly=3,5,7$ suggest that the corrections to $f(\Ly)$ at finite $\Ly$ are mild, especially considering that similar values of the renormalized coupling $z = \xi_2(\Lx,g)$ were obtained with drastically different lattice spacings (usually $\Lx = 64$ for the smaller $\Ly$ compared to $\Lx = 1024$ for the larger $\Ly$).

  Similar results are also observed with even ladders, where the staggering $\gamma$ is a perturbation about the $\theta=0$ theory.
Preliminary results from $\Ly = 2,4,6$ with alternating and columnar staggerings suggest that $\theta = \pi$ can also be obtained in this way.
These results strongly motivate a conjecture: the continuum limit ($\Ly \to \infty$, with $\Ly$ either odd or even) for each fixed $\gamma \Ly$ is in fact a \emph{unique} \ac{QFT} corresponding to the $1+1$-dimensional $\O3$ \ac{NLSM} with a fixed $\theta$,
\begin{align}
    \theta &\equiv \pi \Ly +  \gamma f  \quad \text{(mod $2\pi$)},
    \label{eq:Conjecture}
\end{align}
where $f$ is a non-universal constant which depends on the details of the model such as choice of couplings, staggering configuration, and whether $\Ly$ is odd or even.
While we have provided strong evidence in favor of this identification, there are many paths forward to establish this more rigorously.
For example, odd and even $\Ly$ could be used to self-validate this conjecture, by showing that their step-scaling functions agree by appropriately tuning $\gamma \Ly$.
Further, a comparison with the approach of Ref.~\cite{bogli_nontrivial_2012} using topological lattice actions would be very illuminating.
In that approach, $\theta$ appears as a manifestly topological parameter and thus does not require an empirical identification like \cref{eq:Conjecture}.
It would also be interesting to connect with analytical results on the $\theta$ vacua based on semiclassical instanton methods.

%%% Local Variables:
%%% mode: latex
%%% TeX-master: "main"
%%% End:

%% file: sec_conclusions.tex
\section{Conclusions}

In this work, we have shown how to implement the $1+1$-dimensional $\O3$ \ac{NLSM} at arbitrary $\theta$ using qubit degrees of freedom.
While the motivation behind this work is the quantum simulation of $\theta$ vacua on near-term quantum hardware, interestingly, this result also advances lattice  computations of \acp{QFT} using \emph{classical} \ac{MC} methods.
On the classical side, it provides the first sign-problem-free \ac{MC} algorithm for arbitrary $\theta$.
Our numerical results, obtained with an efficient worm algorithm, indicate that the entire range of $\theta$ vacua is contained in this model, and we conjecture a simple prescription of how the continuum limit can be reached and the physics at all scales can be studied.

This construction enables real-time simulation of $\theta$ vacua in the $\O3$ \ac{NLSM} on near-term quantum hardware. These theories can be regularized at any lattice spacing through an embedding into a $2$-dimensional square lattice of qubits with nearest-neighbor Heisenberg-type interactions. The alternating staggering is a prime candidate for an analog quantum simulation platform like ultracold atoms, with uniform pairwise interactions, and couplings that can be arranged through the trapping pattern shown in \cref{fig:analog}.
On digital quantum hardware like superconducting qubits or trapped ions, either staggering can be implemented using standard Suzuki-Trotter decompositions.
Interestingly, the limit $\Lx \gg \Ly$ is also amenable to \textsc{dmrg}-type algorithms on tensor networks, which would be powerful complementary approach to lattice \ac{MC} and quantum simulation going forward.

In lattice field theory, the $1+1$-dimensional $\O3$ \ac{NLSM} has been long considered an ideal testbed for static properties of \ac{QCD}, exhibiting many of its features, including asymptotic-freedom and $\theta$ vacua.
Even more possibilities open up once we have access to realtime dynamics using quantum platforms.
For instance, accessing nontrivial $\theta$ would allow the study of inelastic scattering processes in an asymptotically-free theory, which would have been impossible in the integrable $\theta=0,\pi$ theories. 

Formulating \acp{QFT} using qubits can yield unexpected advantages. For the $\O3$ \ac{NLSM}, this approach has the rather remarkable feature that it completely circumvents a sign problem present in conventional lattice formulations of the $\theta$ term, and is amenable to efficient cluster algorithms.
Extension to the entire family of $\CP(N-1)$ models is straightforward.
This is encouraging on the path forward towards studying \ac{QCD} with novel classical and quantum algorithms.
Our results demonstrate that there is no fundamental obstruction to studying $\theta$ vacua with discrete degrees of freedom, but whether such ideas might one day even help with the sign problems in \ac{QCD} remains to be seen.

%%% Local Variables:
%%% mode: latex
%%% TeX-master: "main"
%%% End:

%% file: main.bbl
\begin{thebibliography}{68}%
\makeatletter
\providecommand \@ifxundefined [1]{%
 \@ifx{#1\undefined}
}%
\providecommand \@ifnum [1]{%
 \ifnum #1\expandafter \@firstoftwo
 \else \expandafter \@secondoftwo
 \fi
}%
\providecommand \@ifx [1]{%
 \ifx #1\expandafter \@firstoftwo
 \else \expandafter \@secondoftwo
 \fi
}%
\providecommand \natexlab [1]{#1}%
\providecommand \enquote  [1]{``#1''}%
\providecommand \bibnamefont  [1]{#1}%
\providecommand \bibfnamefont [1]{#1}%
\providecommand \citenamefont [1]{#1}%
\providecommand \href@noop [0]{\@secondoftwo}%
\providecommand \href [0]{\begingroup \@sanitize@url \@href}%
\providecommand \@href[1]{\@@startlink{#1}\@@href}%
\providecommand \@@href[1]{\endgroup#1\@@endlink}%
\providecommand \@sanitize@url [0]{\catcode `\\12\catcode `\$12\catcode
  `\&12\catcode `\#12\catcode `\^12\catcode `\_12\catcode `\%12\relax}%
\providecommand \@@startlink[1]{}%
\providecommand \@@endlink[0]{}%
\providecommand \url  [0]{\begingroup\@sanitize@url \@url }%
\providecommand \@url [1]{\endgroup\@href {#1}{\urlprefix }}%
\providecommand \urlprefix  [0]{URL }%
\providecommand \Eprint [0]{\href }%
\providecommand \doibase [0]{https://doi.org/}%
\providecommand \selectlanguage [0]{\@gobble}%
\providecommand \bibinfo  [0]{\@secondoftwo}%
\providecommand \bibfield  [0]{\@secondoftwo}%
\providecommand \translation [1]{[#1]}%
\providecommand \BibitemOpen [0]{}%
\providecommand \bibitemStop [0]{}%
\providecommand \bibitemNoStop [0]{.\EOS\space}%
\providecommand \EOS [0]{\spacefactor3000\relax}%
\providecommand \BibitemShut  [1]{\csname bibitem#1\endcsname}%
\let\auto@bib@innerbib\@empty
%</preamble>
\bibitem [{\citenamefont {Ratti}(2018)}]{ratti_lattice_2018}%
  \BibitemOpen
  \bibfield  {author} {\bibinfo {author} {\bibfnamefont {C.}~\bibnamefont
  {Ratti}},\ }\bibfield  {title} {\bibinfo {title} {Lattice {{QCD}} and heavy
  ion collisions: A review of recent progress},\ }\href
  {https://doi.org/10.1088/1361-6633/aabb97} {\bibfield  {journal} {\bibinfo
  {journal} {Reports on Progress in Physics}\ }\textbf {\bibinfo {volume}
  {81}},\ \bibinfo {pages} {084301} (\bibinfo {year} {2018})}\BibitemShut
  {NoStop}%
\bibitem [{\citenamefont {Davoudi}\ \emph {et~al.}(2021)\citenamefont
  {Davoudi}, \citenamefont {Detmold}, \citenamefont {Shanahan}, \citenamefont
  {Orginos}, \citenamefont {Parre{\~n}o}, \citenamefont {Savage},\ and\
  \citenamefont {Wagman}}]{davoudi_nuclear_2021}%
  \BibitemOpen
  \bibfield  {author} {\bibinfo {author} {\bibfnamefont {Z.}~\bibnamefont
  {Davoudi}}, \bibinfo {author} {\bibfnamefont {W.}~\bibnamefont {Detmold}},
  \bibinfo {author} {\bibfnamefont {P.}~\bibnamefont {Shanahan}}, \bibinfo
  {author} {\bibfnamefont {K.}~\bibnamefont {Orginos}}, \bibinfo {author}
  {\bibfnamefont {A.}~\bibnamefont {Parre{\~n}o}}, \bibinfo {author}
  {\bibfnamefont {M.~J.}\ \bibnamefont {Savage}},\ and\ \bibinfo {author}
  {\bibfnamefont {M.~L.}\ \bibnamefont {Wagman}},\ }\bibfield  {title}
  {\bibinfo {title} {Nuclear matrix elements from lattice {{QCD}} for
  electroweak and beyond-{{Standard-Model}} processes},\ }\href
  {https://doi.org/10.1016/j.physrep.2020.10.004} {\bibfield  {journal}
  {\bibinfo  {journal} {Physics Reports}\ }\bibinfo {series} {Nuclear Matrix
  Elements from Lattice {{QCD}} for Electroweak and
  beyond\textendash{{Standard-Model}} Processes},\ \textbf {\bibinfo {volume}
  {900}},\ \bibinfo {pages} {1} (\bibinfo {year} {2021})}\BibitemShut {NoStop}%
\bibitem [{\citenamefont {Hansen}\ and\ \citenamefont
  {Sharpe}(2019)}]{hansen_lattice_2019}%
  \BibitemOpen
  \bibfield  {author} {\bibinfo {author} {\bibfnamefont {M.~T.}\ \bibnamefont
  {Hansen}}\ and\ \bibinfo {author} {\bibfnamefont {S.~R.}\ \bibnamefont
  {Sharpe}},\ }\bibfield  {title} {\bibinfo {title} {Lattice {{QCD}} and
  {{Three-Particle Decays}} of {{Resonances}}},\ }\href
  {https://doi.org/10.1146/annurev-nucl-101918-023723} {\bibfield  {journal}
  {\bibinfo  {journal} {Annual Review of Nuclear and Particle Science}\
  }\textbf {\bibinfo {volume} {69}},\ \bibinfo {pages} {65} (\bibinfo {year}
  {2019})}\BibitemShut {NoStop}%
\bibitem [{\citenamefont {Constantinou}\ \emph {et~al.}(2021)\citenamefont
  {Constantinou}, \citenamefont {Courtoy}, \citenamefont {Ebert}, \citenamefont
  {Engelhardt}, \citenamefont {Giani}, \citenamefont {Hobbs}, \citenamefont
  {Hou}, \citenamefont {Kusina}, \citenamefont {Kutak}, \citenamefont {Liang},
  \citenamefont {Lin}, \citenamefont {Liu}, \citenamefont {Liuti},
  \citenamefont {Mezrag}, \citenamefont {Nadolsky}, \citenamefont {Nocera},
  \citenamefont {Olness}, \citenamefont {Qiu}, \citenamefont {Radici},
  \citenamefont {Radyushkin}, \citenamefont {Rajan}, \citenamefont {Rogers},
  \citenamefont {Rojo}, \citenamefont {Schierholz}, \citenamefont {Yuan},
  \citenamefont {Zhang},\ and\ \citenamefont
  {Zhang}}]{constantinou_parton_2021}%
  \BibitemOpen
  \bibfield  {author} {\bibinfo {author} {\bibfnamefont {M.}~\bibnamefont
  {Constantinou}}, \bibinfo {author} {\bibfnamefont {A.}~\bibnamefont
  {Courtoy}}, \bibinfo {author} {\bibfnamefont {M.~A.}\ \bibnamefont {Ebert}},
  \bibinfo {author} {\bibfnamefont {M.}~\bibnamefont {Engelhardt}}, \bibinfo
  {author} {\bibfnamefont {T.}~\bibnamefont {Giani}}, \bibinfo {author}
  {\bibfnamefont {T.}~\bibnamefont {Hobbs}}, \bibinfo {author} {\bibfnamefont
  {T.-J.}\ \bibnamefont {Hou}}, \bibinfo {author} {\bibfnamefont
  {A.}~\bibnamefont {Kusina}}, \bibinfo {author} {\bibfnamefont
  {K.}~\bibnamefont {Kutak}}, \bibinfo {author} {\bibfnamefont
  {J.}~\bibnamefont {Liang}}, \bibinfo {author} {\bibfnamefont {H.-W.}\
  \bibnamefont {Lin}}, \bibinfo {author} {\bibfnamefont {K.-F.}\ \bibnamefont
  {Liu}}, \bibinfo {author} {\bibfnamefont {S.}~\bibnamefont {Liuti}}, \bibinfo
  {author} {\bibfnamefont {C.}~\bibnamefont {Mezrag}}, \bibinfo {author}
  {\bibfnamefont {P.}~\bibnamefont {Nadolsky}}, \bibinfo {author}
  {\bibfnamefont {E.~R.}\ \bibnamefont {Nocera}}, \bibinfo {author}
  {\bibfnamefont {F.}~\bibnamefont {Olness}}, \bibinfo {author} {\bibfnamefont
  {J.-W.}\ \bibnamefont {Qiu}}, \bibinfo {author} {\bibfnamefont
  {M.}~\bibnamefont {Radici}}, \bibinfo {author} {\bibfnamefont
  {A.}~\bibnamefont {Radyushkin}}, \bibinfo {author} {\bibfnamefont
  {A.}~\bibnamefont {Rajan}}, \bibinfo {author} {\bibfnamefont
  {T.}~\bibnamefont {Rogers}}, \bibinfo {author} {\bibfnamefont
  {J.}~\bibnamefont {Rojo}}, \bibinfo {author} {\bibfnamefont {G.}~\bibnamefont
  {Schierholz}}, \bibinfo {author} {\bibfnamefont {C.~P.}\ \bibnamefont
  {Yuan}}, \bibinfo {author} {\bibfnamefont {J.-H.}\ \bibnamefont {Zhang}},\
  and\ \bibinfo {author} {\bibfnamefont {R.}~\bibnamefont {Zhang}},\ }\bibfield
   {title} {\bibinfo {title} {Parton distributions and lattice-{{QCD}}
  calculations: {{Toward 3D}} structure},\ }\href
  {https://doi.org/10.1016/j.ppnp.2021.103908} {\bibfield  {journal} {\bibinfo
  {journal} {Progress in Particle and Nuclear Physics}\ }\textbf {\bibinfo
  {volume} {121}},\ \bibinfo {pages} {103908} (\bibinfo {year}
  {2021})}\BibitemShut {NoStop}%
\bibitem [{\citenamefont {Hsu}\ and\ \citenamefont {Reeb}(2010)}]{hsu2010sign}%
  \BibitemOpen
  \bibfield  {author} {\bibinfo {author} {\bibfnamefont {S.~D.}\ \bibnamefont
  {Hsu}}\ and\ \bibinfo {author} {\bibfnamefont {D.}~\bibnamefont {Reeb}},\
  }\bibfield  {title} {\bibinfo {title} {On the sign problem in dense qcd},\
  }\href@noop {} {\bibfield  {journal} {\bibinfo  {journal} {International
  Journal of Modern Physics A}\ }\textbf {\bibinfo {volume} {25}},\ \bibinfo
  {pages} {53} (\bibinfo {year} {2010})}\BibitemShut {NoStop}%
\bibitem [{\citenamefont {Goy}\ \emph {et~al.}(2017)\citenamefont {Goy},
  \citenamefont {Bornyakov}, \citenamefont {Boyda}, \citenamefont {Molochkov},
  \citenamefont {Nakamura}, \citenamefont {Nikolaev},\ and\ \citenamefont
  {Zakharov}}]{goy2017sign}%
  \BibitemOpen
  \bibfield  {author} {\bibinfo {author} {\bibfnamefont {V.}~\bibnamefont
  {Goy}}, \bibinfo {author} {\bibfnamefont {V.}~\bibnamefont {Bornyakov}},
  \bibinfo {author} {\bibfnamefont {D.}~\bibnamefont {Boyda}}, \bibinfo
  {author} {\bibfnamefont {A.}~\bibnamefont {Molochkov}}, \bibinfo {author}
  {\bibfnamefont {A.}~\bibnamefont {Nakamura}}, \bibinfo {author}
  {\bibfnamefont {A.}~\bibnamefont {Nikolaev}},\ and\ \bibinfo {author}
  {\bibfnamefont {V.}~\bibnamefont {Zakharov}},\ }\bibfield  {title} {\bibinfo
  {title} {Sign problem in finite density lattice qcd},\ }\href@noop {}
  {\bibfield  {journal} {\bibinfo  {journal} {Progress of Theoretical and
  Experimental Physics}\ }\textbf {\bibinfo {volume} {2017}},\ \bibinfo {pages}
  {031D01} (\bibinfo {year} {2017})}\BibitemShut {NoStop}%
\bibitem [{\citenamefont {Jordan}\ \emph {et~al.}(2011)\citenamefont {Jordan},
  \citenamefont {Lee},\ and\ \citenamefont {Preskill}}]{jordan2011quantum}%
  \BibitemOpen
  \bibfield  {author} {\bibinfo {author} {\bibfnamefont {S.~P.}\ \bibnamefont
  {Jordan}}, \bibinfo {author} {\bibfnamefont {K.~S.}\ \bibnamefont {Lee}},\
  and\ \bibinfo {author} {\bibfnamefont {J.}~\bibnamefont {Preskill}},\
  }\bibfield  {title} {\bibinfo {title} {Quantum computation of scattering in
  scalar quantum field theories},\ }\href@noop {} {\bibfield  {journal}
  {\bibinfo  {journal} {arXiv preprint arXiv:1112.4833}\ } (\bibinfo {year}
  {2011})}\BibitemShut {NoStop}%
\bibitem [{\citenamefont {Jordan}\ \emph {et~al.}(2012)\citenamefont {Jordan},
  \citenamefont {Lee},\ and\ \citenamefont {Preskill}}]{jordan2012quantum}%
  \BibitemOpen
  \bibfield  {author} {\bibinfo {author} {\bibfnamefont {S.~P.}\ \bibnamefont
  {Jordan}}, \bibinfo {author} {\bibfnamefont {K.~S.}\ \bibnamefont {Lee}},\
  and\ \bibinfo {author} {\bibfnamefont {J.}~\bibnamefont {Preskill}},\
  }\bibfield  {title} {\bibinfo {title} {Quantum algorithms for quantum field
  theories},\ }\href@noop {} {\bibfield  {journal} {\bibinfo  {journal}
  {Science}\ }\textbf {\bibinfo {volume} {336}},\ \bibinfo {pages} {1130}
  (\bibinfo {year} {2012})}\BibitemShut {NoStop}%
\bibitem [{\citenamefont {Yeter-Aydeniz}\ \emph {et~al.}(2019)\citenamefont
  {Yeter-Aydeniz}, \citenamefont {Dumitrescu}, \citenamefont {McCaskey},
  \citenamefont {Bennink}, \citenamefont {Pooser},\ and\ \citenamefont
  {Siopsis}}]{yeter2019scalar}%
  \BibitemOpen
  \bibfield  {author} {\bibinfo {author} {\bibfnamefont {K.}~\bibnamefont
  {Yeter-Aydeniz}}, \bibinfo {author} {\bibfnamefont {E.~F.}\ \bibnamefont
  {Dumitrescu}}, \bibinfo {author} {\bibfnamefont {A.~J.}\ \bibnamefont
  {McCaskey}}, \bibinfo {author} {\bibfnamefont {R.~S.}\ \bibnamefont
  {Bennink}}, \bibinfo {author} {\bibfnamefont {R.~C.}\ \bibnamefont
  {Pooser}},\ and\ \bibinfo {author} {\bibfnamefont {G.}~\bibnamefont
  {Siopsis}},\ }\bibfield  {title} {\bibinfo {title} {Scalar quantum field
  theories as a benchmark for near-term quantum computers},\ }\href@noop {}
  {\bibfield  {journal} {\bibinfo  {journal} {Physical Review A}\ }\textbf
  {\bibinfo {volume} {99}},\ \bibinfo {pages} {032306} (\bibinfo {year}
  {2019})}\BibitemShut {NoStop}%
\bibitem [{\citenamefont {Klco}\ and\ \citenamefont
  {Savage}(2019)}]{klco2019digitization}%
  \BibitemOpen
  \bibfield  {author} {\bibinfo {author} {\bibfnamefont {N.}~\bibnamefont
  {Klco}}\ and\ \bibinfo {author} {\bibfnamefont {M.~J.}\ \bibnamefont
  {Savage}},\ }\bibfield  {title} {\bibinfo {title} {Digitization of scalar
  fields for quantum computing},\ }\href@noop {} {\bibfield  {journal}
  {\bibinfo  {journal} {Physical Review A}\ }\textbf {\bibinfo {volume} {99}},\
  \bibinfo {pages} {052335} (\bibinfo {year} {2019})}\BibitemShut {NoStop}%
\bibitem [{\citenamefont {Chandrasekharan}\ \emph {et~al.}(2002)\citenamefont
  {Chandrasekharan}, \citenamefont {Scarlet},\ and\ \citenamefont
  {Wiese}}]{chandrasekharan_spin_2002}%
  \BibitemOpen
  \bibfield  {author} {\bibinfo {author} {\bibfnamefont {S.}~\bibnamefont
  {Chandrasekharan}}, \bibinfo {author} {\bibfnamefont {B.}~\bibnamefont
  {Scarlet}},\ and\ \bibinfo {author} {\bibfnamefont {U.~J.}\ \bibnamefont
  {Wiese}},\ }\bibfield  {title} {\bibinfo {title} {From spin ladders to the
  {{2D O}}(3) model at non-zero density},\ }\href
  {https://doi.org/10.1016/S0010-4655(02)00311-9} {\bibfield  {journal}
  {\bibinfo  {journal} {Computer Physics Communications}\ }\bibinfo {series}
  {Proceedings of the {{Europhysics Conference}} on {{Computational Physics
  Computational Modeling}} and {{Simulation}} of {{Complex Systems}}},\ \textbf
  {\bibinfo {volume} {147}},\ \bibinfo {pages} {388} (\bibinfo {year}
  {2002})}\BibitemShut {NoStop}%
\bibitem [{\citenamefont {Brower}\ \emph {et~al.}(2004)\citenamefont {Brower},
  \citenamefont {Chandrasekharan}, \citenamefont {Riederer},\ and\
  \citenamefont {Wiese}}]{brower_dtheory_2004}%
  \BibitemOpen
  \bibfield  {author} {\bibinfo {author} {\bibfnamefont {R.}~\bibnamefont
  {Brower}}, \bibinfo {author} {\bibfnamefont {S.}~\bibnamefont
  {Chandrasekharan}}, \bibinfo {author} {\bibfnamefont {S.}~\bibnamefont
  {Riederer}},\ and\ \bibinfo {author} {\bibfnamefont {U.~J.}\ \bibnamefont
  {Wiese}},\ }\bibfield  {title} {\bibinfo {title} {D-theory: Field
  quantization by dimensional reduction of discrete variables},\ }\href
  {https://doi.org/10.1016/j.nuclphysb.2004.06.007} {\bibfield  {journal}
  {\bibinfo  {journal} {Nuclear Physics B}\ }\textbf {\bibinfo {volume}
  {693}},\ \bibinfo {pages} {149} (\bibinfo {year} {2004})}\BibitemShut
  {NoStop}%
\bibitem [{\citenamefont {Beard}\ \emph {et~al.}(2006)\citenamefont {Beard},
  \citenamefont {Pepe}, \citenamefont {Riederer},\ and\ \citenamefont
  {Wiese}}]{beard_efficient_2006}%
  \BibitemOpen
  \bibfield  {author} {\bibinfo {author} {\bibfnamefont {B.~B.}\ \bibnamefont
  {Beard}}, \bibinfo {author} {\bibfnamefont {M.}~\bibnamefont {Pepe}},
  \bibinfo {author} {\bibfnamefont {S.}~\bibnamefont {Riederer}},\ and\
  \bibinfo {author} {\bibfnamefont {U.-J.}\ \bibnamefont {Wiese}},\ }\bibfield
  {title} {\bibinfo {title} {Efficient {{Cluster Algorithm}} for
  {{CP}}({{N-1}}) {{Models}}},\ }\href
  {https://doi.org/10.1016/j.cpc.2006.06.007} {\bibfield  {journal} {\bibinfo
  {journal} {Computer Physics Communications}\ }\textbf {\bibinfo {volume}
  {175}},\ \bibinfo {pages} {629} (\bibinfo {year} {2006})}\BibitemShut
  {NoStop}%
\bibitem [{\citenamefont {Laflamme}\ \emph {et~al.}(2016)\citenamefont
  {Laflamme}, \citenamefont {Evans}, \citenamefont {Dalmonte}, \citenamefont
  {Gerber}, \citenamefont {{Mej{\'i}a-D{\'i}az}}, \citenamefont {Bietenholz},
  \citenamefont {Wiese},\ and\ \citenamefont {Zoller}}]{laflamme_cp_2016}%
  \BibitemOpen
  \bibfield  {author} {\bibinfo {author} {\bibfnamefont {C.}~\bibnamefont
  {Laflamme}}, \bibinfo {author} {\bibfnamefont {W.}~\bibnamefont {Evans}},
  \bibinfo {author} {\bibfnamefont {M.}~\bibnamefont {Dalmonte}}, \bibinfo
  {author} {\bibfnamefont {U.}~\bibnamefont {Gerber}}, \bibinfo {author}
  {\bibfnamefont {H.}~\bibnamefont {{Mej{\'i}a-D{\'i}az}}}, \bibinfo {author}
  {\bibfnamefont {W.}~\bibnamefont {Bietenholz}}, \bibinfo {author}
  {\bibfnamefont {U.~J.}\ \bibnamefont {Wiese}},\ and\ \bibinfo {author}
  {\bibfnamefont {P.}~\bibnamefont {Zoller}},\ }\bibfield  {title} {\bibinfo
  {title} {{{CP}}({{N}}-1) ~quantum field theories with alkaline-earth atoms in
  optical lattices},\ }\href {https://doi.org/10.1016/j.aop.2016.03.012}
  {\bibfield  {journal} {\bibinfo  {journal} {Annals of Physics}\ }\textbf
  {\bibinfo {volume} {370}},\ \bibinfo {pages} {117} (\bibinfo {year}
  {2016})}\BibitemShut {NoStop}%
\bibitem [{\citenamefont {Evans}\ \emph {et~al.}(2018)\citenamefont {Evans},
  \citenamefont {Gerber}, \citenamefont {Hornung},\ and\ \citenamefont
  {Wiese}}]{evans_su_2018a}%
  \BibitemOpen
  \bibfield  {author} {\bibinfo {author} {\bibfnamefont {W.}~\bibnamefont
  {Evans}}, \bibinfo {author} {\bibfnamefont {U.}~\bibnamefont {Gerber}},
  \bibinfo {author} {\bibfnamefont {M.}~\bibnamefont {Hornung}},\ and\ \bibinfo
  {author} {\bibfnamefont {U.~J.}\ \bibnamefont {Wiese}},\ }\bibfield  {title}
  {\bibinfo {title} {{{SU}}(3) quantum spin ladders as a regularization of the
  {{CP}}(2) model at non-zero density: {{From}} classical to quantum
  simulation},\ }\href {https://doi.org/10.1016/j.aop.2018.09.002} {\bibfield
  {journal} {\bibinfo  {journal} {Annals of Physics}\ }\textbf {\bibinfo
  {volume} {398}},\ \bibinfo {pages} {94} (\bibinfo {year} {2018})}\BibitemShut
  {NoStop}%
\bibitem [{\citenamefont {Bruckmann}\ \emph {et~al.}(2019)\citenamefont
  {Bruckmann}, \citenamefont {Jansen},\ and\ \citenamefont
  {K{\"u}hn}}]{bruckmann_nonlinear_2019}%
  \BibitemOpen
  \bibfield  {author} {\bibinfo {author} {\bibfnamefont {F.}~\bibnamefont
  {Bruckmann}}, \bibinfo {author} {\bibfnamefont {K.}~\bibnamefont {Jansen}},\
  and\ \bibinfo {author} {\bibfnamefont {S.}~\bibnamefont {K{\"u}hn}},\
  }\bibfield  {title} {\bibinfo {title} {O(3) nonlinear sigma model in \$1+1\$
  dimensions with matrix product states},\ }\href
  {https://doi.org/10.1103/PhysRevD.99.074501} {\bibfield  {journal} {\bibinfo
  {journal} {Physical Review D}\ }\textbf {\bibinfo {volume} {99}},\ \bibinfo
  {pages} {074501} (\bibinfo {year} {2019})}\BibitemShut {NoStop}%
\bibitem [{\citenamefont {Singh}\ and\ \citenamefont
  {Chandrasekharan}(2019)}]{singh_qubit_2019}%
  \BibitemOpen
  \bibfield  {author} {\bibinfo {author} {\bibfnamefont {H.}~\bibnamefont
  {Singh}}\ and\ \bibinfo {author} {\bibfnamefont {S.}~\bibnamefont
  {Chandrasekharan}},\ }\bibfield  {title} {\bibinfo {title} {A qubit
  regularization of the ${O(3)}$ sigma model},\ }\href
  {https://doi.org/10.1103/PhysRevD.100.054505} {\bibfield  {journal} {\bibinfo
   {journal} {Phys.Rev.}\ }\textbf {\bibinfo {volume} {D100}},\ \bibinfo
  {pages} {054505} (\bibinfo {year} {2019})}\BibitemShut {NoStop}%
\bibitem [{\citenamefont {Bhattacharya}\ \emph {et~al.}(2021)\citenamefont
  {Bhattacharya}, \citenamefont {Buser}, \citenamefont {Chandrasekharan},
  \citenamefont {Gupta},\ and\ \citenamefont
  {Singh}}]{bhattacharya_qubit_2021}%
  \BibitemOpen
  \bibfield  {author} {\bibinfo {author} {\bibfnamefont {T.}~\bibnamefont
  {Bhattacharya}}, \bibinfo {author} {\bibfnamefont {A.~J.}\ \bibnamefont
  {Buser}}, \bibinfo {author} {\bibfnamefont {S.}~\bibnamefont
  {Chandrasekharan}}, \bibinfo {author} {\bibfnamefont {R.}~\bibnamefont
  {Gupta}},\ and\ \bibinfo {author} {\bibfnamefont {H.}~\bibnamefont {Singh}},\
  }\bibfield  {title} {\bibinfo {title} {Qubit regularization of asymptotic
  freedom},\ }\href {https://doi.org/10.1103/PhysRevLett.126.172001} {\bibfield
   {journal} {\bibinfo  {journal} {Physical Review Letters}\ }\textbf {\bibinfo
  {volume} {126}},\ \bibinfo {pages} {172001} (\bibinfo {year}
  {2021})}\BibitemShut {NoStop}%
\bibitem [{\citenamefont {Singh}(2019)}]{singh_qubit_2019a}%
  \BibitemOpen
  \bibfield  {author} {\bibinfo {author} {\bibfnamefont {H.}~\bibnamefont
  {Singh}},\ }\bibfield  {title} {\bibinfo {title} {Qubit $o(n)$ nonlinear
  sigma models},\ }\href@noop {} {\bibfield  {journal} {\bibinfo  {journal}
  {arXiv:1911.12353 [hep-lat, physics:quant-ph]}\ } (\bibinfo {year}
  {2019})}\BibitemShut {NoStop}%
\bibitem [{\citenamefont {Singh}(2022)}]{singh_largecharge_2022a}%
  \BibitemOpen
  \bibfield  {author} {\bibinfo {author} {\bibfnamefont {H.}~\bibnamefont
  {Singh}},\ }\bibfield  {title} {\bibinfo {title} {Large-charge conformal
  dimensions at the \${{O}}({{N}})\$ {{Wilson-Fisher}} fixed point},\
  }\href@noop {} {\bibfield  {journal} {\bibinfo  {journal} {arXiv:2203.00059
  [hep-lat]}\ } (\bibinfo {year} {2022})}\BibitemShut {NoStop}%
\bibitem [{\citenamefont {Chandrasekharan}\ and\ \citenamefont
  {Wiese}(1997)}]{chandrasekharan_quantum_1997a}%
  \BibitemOpen
  \bibfield  {author} {\bibinfo {author} {\bibfnamefont {S.}~\bibnamefont
  {Chandrasekharan}}\ and\ \bibinfo {author} {\bibfnamefont {U.~J.}\
  \bibnamefont {Wiese}},\ }\bibfield  {title} {\bibinfo {title} {Quantum link
  models: {{A}} discrete approach to gauge theories},\ }\href
  {https://doi.org/10.1016/S0550-3213(97)80041-7} {\bibfield  {journal}
  {\bibinfo  {journal} {Nuclear Physics B}\ }\textbf {\bibinfo {volume}
  {492}},\ \bibinfo {pages} {455} (\bibinfo {year} {1997})}\BibitemShut
  {NoStop}%
\bibitem [{\citenamefont {Brower}\ \emph {et~al.}(1999)\citenamefont {Brower},
  \citenamefont {Chandrasekharan},\ and\ \citenamefont
  {Wiese}}]{brower_qcd_1999}%
  \BibitemOpen
  \bibfield  {author} {\bibinfo {author} {\bibfnamefont {R.}~\bibnamefont
  {Brower}}, \bibinfo {author} {\bibfnamefont {S.}~\bibnamefont
  {Chandrasekharan}},\ and\ \bibinfo {author} {\bibfnamefont {U.~J.}\
  \bibnamefont {Wiese}},\ }\bibfield  {title} {\bibinfo {title} {{{QCD}} as a
  quantum link model},\ }\href {https://doi.org/10.1103/PhysRevD.60.094502}
  {\bibfield  {journal} {\bibinfo  {journal} {Phys.Rev.}\ }\textbf {\bibinfo
  {volume} {D60}},\ \bibinfo {pages} {094502} (\bibinfo {year}
  {1999})}\BibitemShut {NoStop}%
\bibitem [{\citenamefont {Raychowdhury}\ and\ \citenamefont
  {Stryker}(2020)}]{raychowdhury2020solving}%
  \BibitemOpen
  \bibfield  {author} {\bibinfo {author} {\bibfnamefont {I.}~\bibnamefont
  {Raychowdhury}}\ and\ \bibinfo {author} {\bibfnamefont {J.~R.}\ \bibnamefont
  {Stryker}},\ }\bibfield  {title} {\bibinfo {title} {Solving gauss's law on
  digital quantum computers with loop-string-hadron digitization},\ }\href@noop
  {} {\bibfield  {journal} {\bibinfo  {journal} {Physical Review Research}\
  }\textbf {\bibinfo {volume} {2}},\ \bibinfo {pages} {033039} (\bibinfo {year}
  {2020})}\BibitemShut {NoStop}%
\bibitem [{\citenamefont {Anishetty}\ \emph {et~al.}(2009)\citenamefont
  {Anishetty}, \citenamefont {Mathur},\ and\ \citenamefont
  {Raychowdhury}}]{anishetty2009prepotential}%
  \BibitemOpen
  \bibfield  {author} {\bibinfo {author} {\bibfnamefont {R.}~\bibnamefont
  {Anishetty}}, \bibinfo {author} {\bibfnamefont {M.}~\bibnamefont {Mathur}},\
  and\ \bibinfo {author} {\bibfnamefont {I.}~\bibnamefont {Raychowdhury}},\
  }\bibfield  {title} {\bibinfo {title} {Prepotential formulation of su (3)
  lattice gauge theory},\ }\href@noop {} {\bibfield  {journal} {\bibinfo
  {journal} {Journal of Physics A: Mathematical and Theoretical}\ }\textbf
  {\bibinfo {volume} {43}},\ \bibinfo {pages} {035403} (\bibinfo {year}
  {2009})}\BibitemShut {NoStop}%
\bibitem [{\citenamefont {Banerjee}\ \emph {et~al.}(2013)\citenamefont
  {Banerjee}, \citenamefont {B{\"o}gli}, \citenamefont {Dalmonte},
  \citenamefont {Rico}, \citenamefont {Stebler}, \citenamefont {Wiese},\ and\
  \citenamefont {Zoller}}]{banerjee2013atomic}%
  \BibitemOpen
  \bibfield  {author} {\bibinfo {author} {\bibfnamefont {D.}~\bibnamefont
  {Banerjee}}, \bibinfo {author} {\bibfnamefont {M.}~\bibnamefont {B{\"o}gli}},
  \bibinfo {author} {\bibfnamefont {M.}~\bibnamefont {Dalmonte}}, \bibinfo
  {author} {\bibfnamefont {E.}~\bibnamefont {Rico}}, \bibinfo {author}
  {\bibfnamefont {P.}~\bibnamefont {Stebler}}, \bibinfo {author} {\bibfnamefont
  {U.-J.}\ \bibnamefont {Wiese}},\ and\ \bibinfo {author} {\bibfnamefont
  {P.}~\bibnamefont {Zoller}},\ }\bibfield  {title} {\bibinfo {title} {Atomic
  quantum simulation of u (n) and su (n) non-abelian lattice gauge theories},\
  }\href@noop {} {\bibfield  {journal} {\bibinfo  {journal} {Physical review
  letters}\ }\textbf {\bibinfo {volume} {110}},\ \bibinfo {pages} {125303}
  (\bibinfo {year} {2013})}\BibitemShut {NoStop}%
\bibitem [{\citenamefont {Zohar}\ \emph {et~al.}(2015)\citenamefont {Zohar},
  \citenamefont {Cirac},\ and\ \citenamefont {Reznik}}]{zohar2015quantum}%
  \BibitemOpen
  \bibfield  {author} {\bibinfo {author} {\bibfnamefont {E.}~\bibnamefont
  {Zohar}}, \bibinfo {author} {\bibfnamefont {J.~I.}\ \bibnamefont {Cirac}},\
  and\ \bibinfo {author} {\bibfnamefont {B.}~\bibnamefont {Reznik}},\
  }\bibfield  {title} {\bibinfo {title} {Quantum simulations of lattice gauge
  theories using ultracold atoms in optical lattices},\ }\href@noop {}
  {\bibfield  {journal} {\bibinfo  {journal} {Reports on Progress in Physics}\
  }\textbf {\bibinfo {volume} {79}},\ \bibinfo {pages} {014401} (\bibinfo
  {year} {2015})}\BibitemShut {NoStop}%
\bibitem [{\citenamefont {Banuls}\ \emph {et~al.}(2017)\citenamefont {Banuls},
  \citenamefont {Cichy}, \citenamefont {Cirac}, \citenamefont {Jansen},\ and\
  \citenamefont {K{\"u}hn}}]{banuls2017efficient}%
  \BibitemOpen
  \bibfield  {author} {\bibinfo {author} {\bibfnamefont {M.~C.}\ \bibnamefont
  {Banuls}}, \bibinfo {author} {\bibfnamefont {K.}~\bibnamefont {Cichy}},
  \bibinfo {author} {\bibfnamefont {J.~I.}\ \bibnamefont {Cirac}}, \bibinfo
  {author} {\bibfnamefont {K.}~\bibnamefont {Jansen}},\ and\ \bibinfo {author}
  {\bibfnamefont {S.}~\bibnamefont {K{\"u}hn}},\ }\bibfield  {title} {\bibinfo
  {title} {Efficient basis formulation for (1+ 1)-dimensional su (2) lattice
  gauge theory: spectral calculations with matrix product states},\ }\href@noop
  {} {\bibfield  {journal} {\bibinfo  {journal} {Physical Review X}\ }\textbf
  {\bibinfo {volume} {7}},\ \bibinfo {pages} {041046} (\bibinfo {year}
  {2017})}\BibitemShut {NoStop}%
\bibitem [{\citenamefont {Muschik}\ \emph {et~al.}(2017)\citenamefont
  {Muschik}, \citenamefont {Heyl}, \citenamefont {Martinez}, \citenamefont
  {Monz}, \citenamefont {Schindler}, \citenamefont {Vogell}, \citenamefont
  {Dalmonte}, \citenamefont {Hauke}, \citenamefont {Blatt},\ and\ \citenamefont
  {Zoller}}]{muschik2017u}%
  \BibitemOpen
  \bibfield  {author} {\bibinfo {author} {\bibfnamefont {C.}~\bibnamefont
  {Muschik}}, \bibinfo {author} {\bibfnamefont {M.}~\bibnamefont {Heyl}},
  \bibinfo {author} {\bibfnamefont {E.}~\bibnamefont {Martinez}}, \bibinfo
  {author} {\bibfnamefont {T.}~\bibnamefont {Monz}}, \bibinfo {author}
  {\bibfnamefont {P.}~\bibnamefont {Schindler}}, \bibinfo {author}
  {\bibfnamefont {B.}~\bibnamefont {Vogell}}, \bibinfo {author} {\bibfnamefont
  {M.}~\bibnamefont {Dalmonte}}, \bibinfo {author} {\bibfnamefont
  {P.}~\bibnamefont {Hauke}}, \bibinfo {author} {\bibfnamefont
  {R.}~\bibnamefont {Blatt}},\ and\ \bibinfo {author} {\bibfnamefont
  {P.}~\bibnamefont {Zoller}},\ }\bibfield  {title} {\bibinfo {title} {U (1)
  wilson lattice gauge theories in digital quantum simulators},\ }\href@noop {}
  {\bibfield  {journal} {\bibinfo  {journal} {New Journal of Physics}\ }\textbf
  {\bibinfo {volume} {19}},\ \bibinfo {pages} {103020} (\bibinfo {year}
  {2017})}\BibitemShut {NoStop}%
\bibitem [{\citenamefont {Zache}\ \emph {et~al.}(2018)\citenamefont {Zache},
  \citenamefont {Hebenstreit}, \citenamefont {Jendrzejewski}, \citenamefont
  {Oberthaler}, \citenamefont {Berges},\ and\ \citenamefont
  {Hauke}}]{zache2018quantum}%
  \BibitemOpen
  \bibfield  {author} {\bibinfo {author} {\bibfnamefont {T.~V.}\ \bibnamefont
  {Zache}}, \bibinfo {author} {\bibfnamefont {F.}~\bibnamefont {Hebenstreit}},
  \bibinfo {author} {\bibfnamefont {F.}~\bibnamefont {Jendrzejewski}}, \bibinfo
  {author} {\bibfnamefont {M.}~\bibnamefont {Oberthaler}}, \bibinfo {author}
  {\bibfnamefont {J.}~\bibnamefont {Berges}},\ and\ \bibinfo {author}
  {\bibfnamefont {P.}~\bibnamefont {Hauke}},\ }\bibfield  {title} {\bibinfo
  {title} {Quantum simulation of lattice gauge theories using wilson
  fermions},\ }\href@noop {} {\bibfield  {journal} {\bibinfo  {journal}
  {Quantum science and technology}\ }\textbf {\bibinfo {volume} {3}},\ \bibinfo
  {pages} {034010} (\bibinfo {year} {2018})}\BibitemShut {NoStop}%
\bibitem [{\citenamefont {Alexandru}\ \emph {et~al.}(2019)\citenamefont
  {Alexandru}, \citenamefont {Bedaque}, \citenamefont {Harmalkar},
  \citenamefont {Lamm}, \citenamefont {Lawrence}, \citenamefont {Warrington},
  \citenamefont {Collaboration} \emph {et~al.}}]{alexandru2019gluon}%
  \BibitemOpen
  \bibfield  {author} {\bibinfo {author} {\bibfnamefont {A.}~\bibnamefont
  {Alexandru}}, \bibinfo {author} {\bibfnamefont {P.~F.}\ \bibnamefont
  {Bedaque}}, \bibinfo {author} {\bibfnamefont {S.}~\bibnamefont {Harmalkar}},
  \bibinfo {author} {\bibfnamefont {H.}~\bibnamefont {Lamm}}, \bibinfo {author}
  {\bibfnamefont {S.}~\bibnamefont {Lawrence}}, \bibinfo {author}
  {\bibfnamefont {N.~C.}\ \bibnamefont {Warrington}}, \bibinfo {author}
  {\bibfnamefont {N.}~\bibnamefont {Collaboration}}, \emph {et~al.},\
  }\bibfield  {title} {\bibinfo {title} {Gluon field digitization for quantum
  computers},\ }\href@noop {} {\bibfield  {journal} {\bibinfo  {journal}
  {Physical Review D}\ }\textbf {\bibinfo {volume} {100}},\ \bibinfo {pages}
  {114501} (\bibinfo {year} {2019})}\BibitemShut {NoStop}%
\bibitem [{\citenamefont {Bender}\ and\ \citenamefont
  {Zohar}(2020)}]{bender2020gauge}%
  \BibitemOpen
  \bibfield  {author} {\bibinfo {author} {\bibfnamefont {J.}~\bibnamefont
  {Bender}}\ and\ \bibinfo {author} {\bibfnamefont {E.}~\bibnamefont {Zohar}},\
  }\bibfield  {title} {\bibinfo {title} {Gauge redundancy-free formulation of
  compact qed with dynamical matter for quantum and classical computations},\
  }\href@noop {} {\bibfield  {journal} {\bibinfo  {journal} {Physical Review
  D}\ }\textbf {\bibinfo {volume} {102}},\ \bibinfo {pages} {114517} (\bibinfo
  {year} {2020})}\BibitemShut {NoStop}%
\bibitem [{\citenamefont {Davoudi}\ \emph {et~al.}(2020)\citenamefont
  {Davoudi}, \citenamefont {Hafezi}, \citenamefont {Monroe}, \citenamefont
  {Pagano}, \citenamefont {Seif},\ and\ \citenamefont
  {Shaw}}]{davoudi2020towards}%
  \BibitemOpen
  \bibfield  {author} {\bibinfo {author} {\bibfnamefont {Z.}~\bibnamefont
  {Davoudi}}, \bibinfo {author} {\bibfnamefont {M.}~\bibnamefont {Hafezi}},
  \bibinfo {author} {\bibfnamefont {C.}~\bibnamefont {Monroe}}, \bibinfo
  {author} {\bibfnamefont {G.}~\bibnamefont {Pagano}}, \bibinfo {author}
  {\bibfnamefont {A.}~\bibnamefont {Seif}},\ and\ \bibinfo {author}
  {\bibfnamefont {A.}~\bibnamefont {Shaw}},\ }\bibfield  {title} {\bibinfo
  {title} {Towards analog quantum simulations of lattice gauge theories with
  trapped ions},\ }\href@noop {} {\bibfield  {journal} {\bibinfo  {journal}
  {Physical Review Research}\ }\textbf {\bibinfo {volume} {2}},\ \bibinfo
  {pages} {023015} (\bibinfo {year} {2020})}\BibitemShut {NoStop}%
\bibitem [{\citenamefont {Klco}\ \emph {et~al.}(2020)\citenamefont {Klco},
  \citenamefont {Savage},\ and\ \citenamefont {Stryker}}]{klco20202}%
  \BibitemOpen
  \bibfield  {author} {\bibinfo {author} {\bibfnamefont {N.}~\bibnamefont
  {Klco}}, \bibinfo {author} {\bibfnamefont {M.~J.}\ \bibnamefont {Savage}},\
  and\ \bibinfo {author} {\bibfnamefont {J.~R.}\ \bibnamefont {Stryker}},\
  }\bibfield  {title} {\bibinfo {title} {Su (2) non-abelian gauge field theory
  in one dimension on digital quantum computers},\ }\href@noop {} {\bibfield
  {journal} {\bibinfo  {journal} {Physical Review D}\ }\textbf {\bibinfo
  {volume} {101}},\ \bibinfo {pages} {074512} (\bibinfo {year}
  {2020})}\BibitemShut {NoStop}%
\bibitem [{\citenamefont {Shaw}\ \emph {et~al.}(2020)\citenamefont {Shaw},
  \citenamefont {Lougovski}, \citenamefont {Stryker},\ and\ \citenamefont
  {Wiebe}}]{shaw2020quantum}%
  \BibitemOpen
  \bibfield  {author} {\bibinfo {author} {\bibfnamefont {A.~F.}\ \bibnamefont
  {Shaw}}, \bibinfo {author} {\bibfnamefont {P.}~\bibnamefont {Lougovski}},
  \bibinfo {author} {\bibfnamefont {J.~R.}\ \bibnamefont {Stryker}},\ and\
  \bibinfo {author} {\bibfnamefont {N.}~\bibnamefont {Wiebe}},\ }\bibfield
  {title} {\bibinfo {title} {Quantum algorithms for simulating the lattice
  schwinger model},\ }\href@noop {} {\bibfield  {journal} {\bibinfo  {journal}
  {Quantum}\ }\textbf {\bibinfo {volume} {4}},\ \bibinfo {pages} {306}
  (\bibinfo {year} {2020})}\BibitemShut {NoStop}%
\bibitem [{\citenamefont {Kasper}\ \emph {et~al.}(2020)\citenamefont {Kasper},
  \citenamefont {Zache}, \citenamefont {Jendrzejewski}, \citenamefont
  {Lewenstein},\ and\ \citenamefont {Zohar}}]{kasper2020non}%
  \BibitemOpen
  \bibfield  {author} {\bibinfo {author} {\bibfnamefont {V.}~\bibnamefont
  {Kasper}}, \bibinfo {author} {\bibfnamefont {T.~V.}\ \bibnamefont {Zache}},
  \bibinfo {author} {\bibfnamefont {F.}~\bibnamefont {Jendrzejewski}}, \bibinfo
  {author} {\bibfnamefont {M.}~\bibnamefont {Lewenstein}},\ and\ \bibinfo
  {author} {\bibfnamefont {E.}~\bibnamefont {Zohar}},\ }\bibfield  {title}
  {\bibinfo {title} {Non-abelian gauge invariance from dynamical decoupling},\
  }\href@noop {} {\bibfield  {journal} {\bibinfo  {journal} {arXiv preprint
  arXiv:2012.08620}\ } (\bibinfo {year} {2020})}\BibitemShut {NoStop}%
\bibitem [{\citenamefont {Buser}\ \emph {et~al.}(2021)\citenamefont {Buser},
  \citenamefont {Gharibyan}, \citenamefont {Hanada}, \citenamefont {Honda},\
  and\ \citenamefont {Liu}}]{buser2021quantum}%
  \BibitemOpen
  \bibfield  {author} {\bibinfo {author} {\bibfnamefont {A.~J.}\ \bibnamefont
  {Buser}}, \bibinfo {author} {\bibfnamefont {H.}~\bibnamefont {Gharibyan}},
  \bibinfo {author} {\bibfnamefont {M.}~\bibnamefont {Hanada}}, \bibinfo
  {author} {\bibfnamefont {M.}~\bibnamefont {Honda}},\ and\ \bibinfo {author}
  {\bibfnamefont {J.}~\bibnamefont {Liu}},\ }\bibfield  {title} {\bibinfo
  {title} {Quantum simulation of gauge theory via orbifold lattice},\
  }\href@noop {} {\bibfield  {journal} {\bibinfo  {journal} {Journal of High
  Energy Physics}\ }\textbf {\bibinfo {volume} {2021}},\ \bibinfo {pages} {1}
  (\bibinfo {year} {2021})}\BibitemShut {NoStop}%
\bibitem [{\citenamefont {Haase}\ \emph {et~al.}(2021)\citenamefont {Haase},
  \citenamefont {Dellantonio}, \citenamefont {Celi}, \citenamefont {Paulson},
  \citenamefont {Kan}, \citenamefont {Jansen},\ and\ \citenamefont
  {Muschik}}]{haase2021resource}%
  \BibitemOpen
  \bibfield  {author} {\bibinfo {author} {\bibfnamefont {J.~F.}\ \bibnamefont
  {Haase}}, \bibinfo {author} {\bibfnamefont {L.}~\bibnamefont {Dellantonio}},
  \bibinfo {author} {\bibfnamefont {A.}~\bibnamefont {Celi}}, \bibinfo {author}
  {\bibfnamefont {D.}~\bibnamefont {Paulson}}, \bibinfo {author} {\bibfnamefont
  {A.}~\bibnamefont {Kan}}, \bibinfo {author} {\bibfnamefont {K.}~\bibnamefont
  {Jansen}},\ and\ \bibinfo {author} {\bibfnamefont {C.~A.}\ \bibnamefont
  {Muschik}},\ }\bibfield  {title} {\bibinfo {title} {A resource efficient
  approach for quantum and classical simulations of gauge theories in particle
  physics},\ }\href@noop {} {\bibfield  {journal} {\bibinfo  {journal}
  {Quantum}\ }\textbf {\bibinfo {volume} {5}},\ \bibinfo {pages} {393}
  (\bibinfo {year} {2021})}\BibitemShut {NoStop}%
\bibitem [{\citenamefont {Zamolodchikov}\ and\ \citenamefont
  {Zamolodchikov}(1979)}]{zamolodchikov_factorized_1979}%
  \BibitemOpen
  \bibfield  {author} {\bibinfo {author} {\bibfnamefont {A.~B.}\ \bibnamefont
  {Zamolodchikov}}\ and\ \bibinfo {author} {\bibfnamefont {A.~B.}\ \bibnamefont
  {Zamolodchikov}},\ }\bibfield  {title} {\bibinfo {title} {Factorized
  {{S-matrices}} in two dimensions as the exact solutions of certain
  relativistic quantum field theory models},\ }\href
  {https://doi.org/10.1016/0003-4916(79)90391-9} {\bibfield  {journal}
  {\bibinfo  {journal} {Annals of Physics}\ }\textbf {\bibinfo {volume}
  {120}},\ \bibinfo {pages} {253} (\bibinfo {year} {1979})}\BibitemShut
  {NoStop}%
\bibitem [{\citenamefont {Balog}\ and\ \citenamefont {Heged{\H
  u}s}(2004)}]{balog_tba_2004}%
  \BibitemOpen
  \bibfield  {author} {\bibinfo {author} {\bibfnamefont {J.}~\bibnamefont
  {Balog}}\ and\ \bibinfo {author} {\bibfnamefont {{\'A}.}~\bibnamefont
  {Heged{\H u}s}},\ }\bibfield  {title} {\bibinfo {title} {{{TBA}} equations
  for excited states in the {{O}}(3) {{andO}}(4) nonlinear -model},\ }\href
  {https://doi.org/10.1088/0305-4470/37/5/027} {\bibfield  {journal} {\bibinfo
  {journal} {Journal of Physics A: Mathematical and General}\ }\textbf
  {\bibinfo {volume} {37}},\ \bibinfo {pages} {1881} (\bibinfo {year}
  {2004})}\BibitemShut {NoStop}%
\bibitem [{\citenamefont {Balog}\ and\ \citenamefont {Heged{\H
  u}s}(2010)}]{balog_finite_2010}%
  \BibitemOpen
  \bibfield  {author} {\bibinfo {author} {\bibfnamefont {J.}~\bibnamefont
  {Balog}}\ and\ \bibinfo {author} {\bibfnamefont {{\'A}.}~\bibnamefont
  {Heged{\H u}s}},\ }\bibfield  {title} {\bibinfo {title} {The finite size
  spectrum of the 2-dimensional {{O}}(3) nonlinear {$\sigma$}-model},\ }\href
  {https://doi.org/10.1016/j.nuclphysb.2009.11.010} {\bibfield  {journal}
  {\bibinfo  {journal} {Nuclear Physics B}\ }\textbf {\bibinfo {volume}
  {829}},\ \bibinfo {pages} {425} (\bibinfo {year} {2010})}\BibitemShut
  {NoStop}%
\bibitem [{\citenamefont {L{\"u}scher}(1978)}]{luscher_quantum_1978}%
  \BibitemOpen
  \bibfield  {author} {\bibinfo {author} {\bibfnamefont {M.}~\bibnamefont
  {L{\"u}scher}},\ }\bibfield  {title} {\bibinfo {title} {Quantum non-local
  charges and absence of particle production in the two-dimensional non-linear
  {$\sigma$}-model},\ }\href {https://doi.org/10.1016/0550-3213(78)90211-0}
  {\bibfield  {journal} {\bibinfo  {journal} {Nuclear Physics B}\ }\textbf
  {\bibinfo {volume} {135}},\ \bibinfo {pages} {1} (\bibinfo {year}
  {1978})}\BibitemShut {NoStop}%
\bibitem [{\citenamefont {L{\"u}scher}\ \emph {et~al.}(1991)\citenamefont
  {L{\"u}scher}, \citenamefont {Weisz},\ and\ \citenamefont
  {Wolff}}]{luscher_numerical_1991}%
  \BibitemOpen
  \bibfield  {author} {\bibinfo {author} {\bibfnamefont {M.}~\bibnamefont
  {L{\"u}scher}}, \bibinfo {author} {\bibfnamefont {P.}~\bibnamefont {Weisz}},\
  and\ \bibinfo {author} {\bibfnamefont {U.}~\bibnamefont {Wolff}},\ }\bibfield
   {title} {\bibinfo {title} {A numerical method to compute the running
  coupling in asymptotically free theories},\ }\href
  {https://doi.org/10.1016/0550-3213(91)90298-C} {\bibfield  {journal}
  {\bibinfo  {journal} {Nuclear Physics B}\ }\textbf {\bibinfo {volume}
  {359}},\ \bibinfo {pages} {221} (\bibinfo {year} {1991})}\BibitemShut
  {NoStop}%
\bibitem [{\citenamefont {Bietenholz}\ \emph {et~al.}(1996)\citenamefont
  {Bietenholz}, \citenamefont {Pochinsky},\ and\ \citenamefont
  {Wiese}}]{bietenholz_testing_1996}%
  \BibitemOpen
  \bibfield  {author} {\bibinfo {author} {\bibfnamefont {W.}~\bibnamefont
  {Bietenholz}}, \bibinfo {author} {\bibfnamefont {A.}~\bibnamefont
  {Pochinsky}},\ and\ \bibinfo {author} {\bibfnamefont {U.-J.}\ \bibnamefont
  {Wiese}},\ }\bibfield  {title} {\bibinfo {title} {Testing {{Haldane}}'s
  {{Conjecture}} in the {{O}}(3) {{Model}} by a {{Meron Cluster Simulation}}},\
  }\href {https://doi.org/10.1016/0920-5632(96)00160-0} {\bibfield  {journal}
  {\bibinfo  {journal} {Nuclear Physics B - Proceedings Supplements}\ }\textbf
  {\bibinfo {volume} {47}},\ \bibinfo {pages} {727} (\bibinfo {year}
  {1996})}\BibitemShut {NoStop}%
\bibitem [{\citenamefont {Caracciolo}\ \emph {et~al.}(1995)\citenamefont
  {Caracciolo}, \citenamefont {Edwards}, \citenamefont {Pelissetto},\ and\
  \citenamefont {Sokal}}]{caracciolo_asymptotic_1995}%
  \BibitemOpen
  \bibfield  {author} {\bibinfo {author} {\bibfnamefont {S.}~\bibnamefont
  {Caracciolo}}, \bibinfo {author} {\bibfnamefont {R.~G.}\ \bibnamefont
  {Edwards}}, \bibinfo {author} {\bibfnamefont {A.}~\bibnamefont
  {Pelissetto}},\ and\ \bibinfo {author} {\bibfnamefont {A.~D.}\ \bibnamefont
  {Sokal}},\ }\bibfield  {title} {\bibinfo {title} {Asymptotic scaling in the
  two-dimensional {O}(3) \ensuremath{\sigma} model at correlation length
  1${0}^{5}$},\ }\href {https://doi.org/10.1103/PhysRevLett.75.1891} {\bibfield
   {journal} {\bibinfo  {journal} {Physical Review Letters}\ }\textbf {\bibinfo
  {volume} {75}},\ \bibinfo {pages} {1891} (\bibinfo {year}
  {1995})}\BibitemShut {NoStop}%
\bibitem [{\citenamefont {Balog}\ \emph {et~al.}(2000)\citenamefont {Balog},
  \citenamefont {Forgacs},\ and\ \citenamefont
  {Palla}}]{balog_twodimensional_2000}%
  \BibitemOpen
  \bibfield  {author} {\bibinfo {author} {\bibfnamefont {J.}~\bibnamefont
  {Balog}}, \bibinfo {author} {\bibfnamefont {P.}~\bibnamefont {Forgacs}},\
  and\ \bibinfo {author} {\bibfnamefont {L.}~\bibnamefont {Palla}},\ }\bibfield
   {title} {\bibinfo {title} {A two-dimensional integrable axionic sigma-model
  and {{T-duality}}},\ }\bibfield  {journal} {\bibinfo  {journal}
  {arXiv:hep-th/0004180}\ }\href
  {https://doi.org/10.1016/S0370-2693(00)00645-6}
  {10.1016/S0370-2693(00)00645-6} (\bibinfo {year} {2000})\BibitemShut
  {NoStop}%
\bibitem [{\citenamefont {Bogli}\ \emph {et~al.}(2012)\citenamefont {Bogli},
  \citenamefont {Niedermayer}, \citenamefont {Pepe},\ and\ \citenamefont
  {Wiese}}]{bogli_nontrivial_2012}%
  \BibitemOpen
  \bibfield  {author} {\bibinfo {author} {\bibfnamefont {M.}~\bibnamefont
  {Bogli}}, \bibinfo {author} {\bibfnamefont {F.}~\bibnamefont {Niedermayer}},
  \bibinfo {author} {\bibfnamefont {M.}~\bibnamefont {Pepe}},\ and\ \bibinfo
  {author} {\bibfnamefont {U.~J.}\ \bibnamefont {Wiese}},\ }\bibfield  {title}
  {\bibinfo {title} {Non-trivial $\theta$-vacuum effects in the 2-d o(3)
  model},\ }\href {https://doi.org/10.1007/JHEP04(2012)117} {\bibfield
  {journal} {\bibinfo  {journal} {JHEP}\ }\textbf {\bibinfo {volume} {04}},\
  \bibinfo {pages} {117}}\BibitemShut {NoStop}%
\bibitem [{\citenamefont {de~Forcrand}\ \emph {et~al.}(2012)\citenamefont
  {de~Forcrand}, \citenamefont {Pepe},\ and\ \citenamefont
  {Wiese}}]{deforcrand_walking_2012}%
  \BibitemOpen
  \bibfield  {author} {\bibinfo {author} {\bibfnamefont {P.}~\bibnamefont
  {de~Forcrand}}, \bibinfo {author} {\bibfnamefont {M.}~\bibnamefont {Pepe}},\
  and\ \bibinfo {author} {\bibfnamefont {U.~J.}\ \bibnamefont {Wiese}},\
  }\bibfield  {title} {\bibinfo {title} {Walking near a conformal fixed point:
  The 2-d $o(3)$ model at
  $\ensuremath{\theta}\mathbf{\ensuremath{\approx}}\ensuremath{\pi}$ as a test
  case},\ }\href {https://doi.org/10.1103/PhysRevD.86.075006} {\bibfield
  {journal} {\bibinfo  {journal} {Physical Review D}\ }\textbf {\bibinfo
  {volume} {86}},\ \bibinfo {pages} {075006} (\bibinfo {year}
  {2012})}\BibitemShut {NoStop}%
\bibitem [{\citenamefont {Wilson}\ and\ \citenamefont
  {Kogut}(1974)}]{wilson_renormalization_1974}%
  \BibitemOpen
  \bibfield  {author} {\bibinfo {author} {\bibfnamefont {K.~G.}\ \bibnamefont
  {Wilson}}\ and\ \bibinfo {author} {\bibfnamefont {J.~B.}\ \bibnamefont
  {Kogut}},\ }\bibfield  {title} {\bibinfo {title} {The {{Renormalization}}
  group and the epsilon expansion},\ }\href
  {https://doi.org/10.1016/0370-1573(74)90023-4} {\bibfield  {journal}
  {\bibinfo  {journal} {Phys.Rept.}\ }\textbf {\bibinfo {volume} {12}},\
  \bibinfo {pages} {75} (\bibinfo {year} {1974})}\BibitemShut {NoStop}%
\bibitem [{\citenamefont {Wilson}(1983)}]{wilson_renormalization_1983}%
  \BibitemOpen
  \bibfield  {author} {\bibinfo {author} {\bibfnamefont {K.~G.}\ \bibnamefont
  {Wilson}},\ }\bibfield  {title} {\bibinfo {title} {The renormalization group
  and critical phenomena},\ }\href {https://doi.org/10.1103/RevModPhys.55.583}
  {\bibfield  {journal} {\bibinfo  {journal} {Reviews of Modern Physics}\
  }\textbf {\bibinfo {volume} {55}},\ \bibinfo {pages} {583} (\bibinfo {year}
  {1983})}\BibitemShut {NoStop}%
\bibitem [{\citenamefont {Kogut}(1979)}]{kogut_introduction_1979}%
  \BibitemOpen
  \bibfield  {author} {\bibinfo {author} {\bibfnamefont {J.~B.}\ \bibnamefont
  {Kogut}},\ }\bibfield  {title} {\bibinfo {title} {An introduction to lattice
  gauge theory and spin systems},\ }\href
  {https://doi.org/10.1103/RevModPhys.51.659} {\bibfield  {journal} {\bibinfo
  {journal} {Reviews of Modern Physics}\ }\textbf {\bibinfo {volume} {51}},\
  \bibinfo {pages} {659} (\bibinfo {year} {1979})}\BibitemShut {NoStop}%
\bibitem [{\citenamefont {Beard}\ \emph {et~al.}(2005)\citenamefont {Beard},
  \citenamefont {Pepe}, \citenamefont {Riederer},\ and\ \citenamefont
  {Wiese}}]{beard_study_2005}%
  \BibitemOpen
  \bibfield  {author} {\bibinfo {author} {\bibfnamefont {B.~B.}\ \bibnamefont
  {Beard}}, \bibinfo {author} {\bibfnamefont {M.}~\bibnamefont {Pepe}},
  \bibinfo {author} {\bibfnamefont {S.}~\bibnamefont {Riederer}},\ and\
  \bibinfo {author} {\bibfnamefont {U.-J.}\ \bibnamefont {Wiese}},\ }\bibfield
  {title} {\bibinfo {title} {Study of $\mathrm{C}\mathrm{P}(n\ensuremath{-}1)$
  $\ensuremath{\theta}$-vacua by cluster simulation of
  $\mathrm{S}\mathrm{U}(n)$ quantum spin ladders},\ }\href
  {https://doi.org/10.1103/PhysRevLett.94.010603} {\bibfield  {journal}
  {\bibinfo  {journal} {Physical Review Letters}\ }\textbf {\bibinfo {volume}
  {94}},\ \bibinfo {pages} {010603} (\bibinfo {year} {2005})}\BibitemShut
  {NoStop}%
\bibitem [{\citenamefont {Caracciolo}\ \emph {et~al.}(1993)\citenamefont
  {Caracciolo}, \citenamefont {Edwards}, \citenamefont {Pelissetto},\ and\
  \citenamefont {Sokal}}]{caracciolo_wolff_1993}%
  \BibitemOpen
  \bibfield  {author} {\bibinfo {author} {\bibfnamefont {S.}~\bibnamefont
  {Caracciolo}}, \bibinfo {author} {\bibfnamefont {R.~G.}\ \bibnamefont
  {Edwards}}, \bibinfo {author} {\bibfnamefont {A.}~\bibnamefont
  {Pelissetto}},\ and\ \bibinfo {author} {\bibfnamefont {A.~D.}\ \bibnamefont
  {Sokal}},\ }\bibfield  {title} {\bibinfo {title} {Wolff type embedding
  algorithms for general nonlinear sigma models},\ }\href
  {https://doi.org/10.1016/0550-3213(93)90044-P} {\bibfield  {journal}
  {\bibinfo  {journal} {Nucl.Phys.}\ }\textbf {\bibinfo {volume} {B403}},\
  \bibinfo {pages} {475} (\bibinfo {year} {1993})}\BibitemShut {NoStop}%
\bibitem [{\citenamefont {Shankar}\ and\ \citenamefont
  {Read}(1990)}]{shankar_th_1990}%
  \BibitemOpen
  \bibfield  {author} {\bibinfo {author} {\bibfnamefont {R.}~\bibnamefont
  {Shankar}}\ and\ \bibinfo {author} {\bibfnamefont {N.}~\bibnamefont {Read}},\
  }\bibfield  {title} {\bibinfo {title} {The \texttheta{} = {$\pi$} nonlinear
  sigma model is massless},\ }\href
  {https://doi.org/10.1016/0550-3213(90)90437-I} {\bibfield  {journal}
  {\bibinfo  {journal} {Nuclear Physics B}\ }\textbf {\bibinfo {volume}
  {336}},\ \bibinfo {pages} {457} (\bibinfo {year} {1990})}\BibitemShut
  {NoStop}%
\bibitem [{\citenamefont {Affleck}(1988)}]{affleck_field_1988}%
  \BibitemOpen
  \bibfield  {author} {\bibinfo {author} {\bibfnamefont {I.}~\bibnamefont
  {Affleck}},\ }\bibfield  {title} {\bibinfo {title} {Field {{Theory Methods}}
  and {{Quantum Critical Phenomena}}},\ }in\ \href@noop {} {\emph {\bibinfo
  {booktitle} {Les {{Houches Summer School}} in {{Theoretical Physics}}:
  {{Fields}}, {{Strings}}, {{Critical Phenomena}}}}}\ (\bibinfo {year}
  {1988})\BibitemShut {NoStop}%
\bibitem [{\citenamefont {{Mart{\'i}n-Delgado}}\ \emph
  {et~al.}(1996)\citenamefont {{Mart{\'i}n-Delgado}}, \citenamefont {Shankar},\
  and\ \citenamefont {Sierra}}]{martin-delgado_phase_1996}%
  \BibitemOpen
  \bibfield  {author} {\bibinfo {author} {\bibfnamefont {M.~A.}\ \bibnamefont
  {{Mart{\'i}n-Delgado}}}, \bibinfo {author} {\bibfnamefont {R.}~\bibnamefont
  {Shankar}},\ and\ \bibinfo {author} {\bibfnamefont {G.}~\bibnamefont
  {Sierra}},\ }\bibfield  {title} {\bibinfo {title} {Phase {{Transitions}} in
  {{Staggered Spin Ladders}}},\ }\href
  {https://doi.org/10.1103/PhysRevLett.77.3443} {\bibfield  {journal} {\bibinfo
   {journal} {Physical Review Letters}\ }\textbf {\bibinfo {volume} {77}},\
  \bibinfo {pages} {3443} (\bibinfo {year} {1996})}\BibitemShut {NoStop}%
\bibitem [{\citenamefont {Sierra}(1996)}]{sierra_nonlinear_1996}%
  \BibitemOpen
  \bibfield  {author} {\bibinfo {author} {\bibfnamefont {G.}~\bibnamefont
  {Sierra}},\ }\bibfield  {title} {\bibinfo {title} {The nonlinear sigma model
  and spin ladders},\ }\href {https://doi.org/10.1088/0305-4470/29/12/032}
  {\bibfield  {journal} {\bibinfo  {journal} {Journal of Physics A:
  Mathematical and General}\ }\textbf {\bibinfo {volume} {29}},\ \bibinfo
  {pages} {3299} (\bibinfo {year} {1996})}\BibitemShut {NoStop}%
\bibitem [{\citenamefont
  {Haldane}(1983{\natexlab{a}})}]{haldane_continuum_1983}%
  \BibitemOpen
  \bibfield  {author} {\bibinfo {author} {\bibfnamefont {F.~D.~M.}\
  \bibnamefont {Haldane}},\ }\bibfield  {title} {\bibinfo {title} {Continuum
  dynamics of the 1-{{D Heisenberg}} antiferromagnetic identification with the
  {{O}}(3) nonlinear sigma model},\ }\href
  {https://doi.org/10.1016/0375-9601(83)90631-X} {\bibfield  {journal}
  {\bibinfo  {journal} {Phys.Lett.}\ }\textbf {\bibinfo {volume} {A93}},\
  \bibinfo {pages} {464} (\bibinfo {year} {1983}{\natexlab{a}})}\BibitemShut
  {NoStop}%
\bibitem [{\citenamefont
  {Haldane}(1983{\natexlab{b}})}]{haldane_nonlinear_1983}%
  \BibitemOpen
  \bibfield  {author} {\bibinfo {author} {\bibfnamefont {F.~D.~M.}\
  \bibnamefont {Haldane}},\ }\bibfield  {title} {\bibinfo {title} {Nonlinear
  field theory of large spin {{Heisenberg}} antiferromagnets.
  {{Semiclassically}} quantized solitons of the one-dimensional easy {{Axis
  Neel}} state},\ }\href {https://doi.org/10.1103/PhysRevLett.50.1153}
  {\bibfield  {journal} {\bibinfo  {journal} {Phys.Rev.Lett.}\ }\textbf
  {\bibinfo {volume} {50}},\ \bibinfo {pages} {1153} (\bibinfo {year}
  {1983}{\natexlab{b}})}\BibitemShut {NoStop}%
\bibitem [{\citenamefont {Affleck}\ and\ \citenamefont
  {Haldane}(1987)}]{affleck_critical_1987}%
  \BibitemOpen
  \bibfield  {author} {\bibinfo {author} {\bibfnamefont {I.}~\bibnamefont
  {Affleck}}\ and\ \bibinfo {author} {\bibfnamefont {F.~D.~M.}\ \bibnamefont
  {Haldane}},\ }\bibfield  {title} {\bibinfo {title} {Critical theory of
  quantum spin chains},\ }\href {https://doi.org/10.1103/PhysRevB.36.5291}
  {\bibfield  {journal} {\bibinfo  {journal} {Physical Review B}\ }\textbf
  {\bibinfo {volume} {36}},\ \bibinfo {pages} {5291} (\bibinfo {year}
  {1987})}\BibitemShut {NoStop}%
\bibitem [{\citenamefont {Venuti}\ \emph {et~al.}(2005)\citenamefont {Venuti},
  \citenamefont {Boschi}, \citenamefont {Ercolessi}, \citenamefont {Ortolani},
  \citenamefont {Morandi}, \citenamefont {Pasini},\ and\ \citenamefont
  {Roncaglia}}]{venuti_particle_2005}%
  \BibitemOpen
  \bibfield  {author} {\bibinfo {author} {\bibfnamefont {L.~C.}\ \bibnamefont
  {Venuti}}, \bibinfo {author} {\bibfnamefont {C.~D.~E.}\ \bibnamefont
  {Boschi}}, \bibinfo {author} {\bibfnamefont {E.}~\bibnamefont {Ercolessi}},
  \bibinfo {author} {\bibfnamefont {F.}~\bibnamefont {Ortolani}}, \bibinfo
  {author} {\bibfnamefont {G.}~\bibnamefont {Morandi}}, \bibinfo {author}
  {\bibfnamefont {S.}~\bibnamefont {Pasini}},\ and\ \bibinfo {author}
  {\bibfnamefont {M.}~\bibnamefont {Roncaglia}},\ }\bibfield  {title} {\bibinfo
  {title} {Particle content of the nonlinear sigma model with a
  \$\textbackslash theta\$-term: A lattice model investigation},\ }\href
  {https://doi.org/10.1088/1742-5468/2005/02/L02004} {\bibfield  {journal}
  {\bibinfo  {journal} {Journal of Statistical Mechanics: Theory and
  Experiment}\ }\textbf {\bibinfo {volume} {2005}},\ \bibinfo {pages} {L02004}
  (\bibinfo {year} {2005})}\BibitemShut {NoStop}%
\bibitem [{\citenamefont {Read}\ and\ \citenamefont
  {Sachdev}(1989{\natexlab{a}})}]{read_valencebond_1989}%
  \BibitemOpen
  \bibfield  {author} {\bibinfo {author} {\bibfnamefont {N.}~\bibnamefont
  {Read}}\ and\ \bibinfo {author} {\bibfnamefont {S.}~\bibnamefont {Sachdev}},\
  }\bibfield  {title} {\bibinfo {title} {Valence-bond and spin-{{Peierls}}
  ground states of low-dimensional quantum antiferromagnets},\ }\href
  {https://doi.org/10.1103/PhysRevLett.62.1694} {\bibfield  {journal} {\bibinfo
   {journal} {Physical Review Letters}\ }\textbf {\bibinfo {volume} {62}},\
  \bibinfo {pages} {1694} (\bibinfo {year} {1989}{\natexlab{a}})}\BibitemShut
  {NoStop}%
\bibitem [{\citenamefont {Read}\ and\ \citenamefont
  {Sachdev}(1989{\natexlab{b}})}]{read_features_1989}%
  \BibitemOpen
  \bibfield  {author} {\bibinfo {author} {\bibfnamefont {N.}~\bibnamefont
  {Read}}\ and\ \bibinfo {author} {\bibfnamefont {S.}~\bibnamefont {Sachdev}},\
  }\bibfield  {title} {\bibinfo {title} {Some features of the phase diagram of
  the square lattice {{SU}}({{N}}) antiferromagnet},\ }\href
  {https://doi.org/10.1016/0550-3213(89)90061-8} {\bibfield  {journal}
  {\bibinfo  {journal} {Nuclear Physics B}\ }\textbf {\bibinfo {volume}
  {316}},\ \bibinfo {pages} {609} (\bibinfo {year}
  {1989}{\natexlab{b}})}\BibitemShut {NoStop}%
\bibitem [{\citenamefont {Prokof'ev}\ and\ \citenamefont
  {Svistunov}(2010)}]{prokofev_worm_2010}%
  \BibitemOpen
  \bibfield  {author} {\bibinfo {author} {\bibfnamefont {N.}~\bibnamefont
  {Prokof'ev}}\ and\ \bibinfo {author} {\bibfnamefont {B.}~\bibnamefont
  {Svistunov}},\ }\bibfield  {title} {\bibinfo {title} {Worm {{Algorithm}} for
  {{Problems}} of {{Quantum}} and {{Classical Statistics}}},\ }\href@noop {}
  {\bibfield  {journal} {\bibinfo  {journal} {arXiv:0910.1393 [cond-mat,
  physics:hep-lat]}\ } (\bibinfo {year} {2010})}\BibitemShut {NoStop}%
\bibitem [{\citenamefont {Prokof'ev}\ and\ \citenamefont
  {Svistunov}(2001)}]{prokofev_worm_2001}%
  \BibitemOpen
  \bibfield  {author} {\bibinfo {author} {\bibfnamefont {N.}~\bibnamefont
  {Prokof'ev}}\ and\ \bibinfo {author} {\bibfnamefont {B.}~\bibnamefont
  {Svistunov}},\ }\bibfield  {title} {\bibinfo {title} {Worm {{Algorithms}} for
  {{Classical Statistical Models}}},\ }\href
  {https://doi.org/10.1103/PhysRevLett.87.160601} {\bibfield  {journal}
  {\bibinfo  {journal} {Phys.Rev.Lett.}\ }\textbf {\bibinfo {volume} {87}},\
  \bibinfo {pages} {160601} (\bibinfo {year} {2001})}\BibitemShut {NoStop}%
\bibitem [{\citenamefont {Wiese}\ and\ \citenamefont
  {Ying}(1994)}]{wiese_determination_1994}%
  \BibitemOpen
  \bibfield  {author} {\bibinfo {author} {\bibfnamefont {U.~J.}\ \bibnamefont
  {Wiese}}\ and\ \bibinfo {author} {\bibfnamefont {H.~P.}\ \bibnamefont
  {Ying}},\ }\bibfield  {title} {\bibinfo {title} {A determination of the low
  energy parameters of the 2-d {{Heisenberg}} antiferromagnet},\ }\href
  {https://doi.org/10.1007/BF01316955} {\bibfield  {journal} {\bibinfo
  {journal} {Zeitschrift f\"ur Physik B Condensed Matter}\ }\textbf {\bibinfo
  {volume} {93}},\ \bibinfo {pages} {147} (\bibinfo {year} {1994})}\BibitemShut
  {NoStop}%
\bibitem [{\citenamefont {Beard}\ and\ \citenamefont
  {Wiese}(1996)}]{beard_simulations_1996}%
  \BibitemOpen
  \bibfield  {author} {\bibinfo {author} {\bibfnamefont {B.~B.}\ \bibnamefont
  {Beard}}\ and\ \bibinfo {author} {\bibfnamefont {U.-J.}\ \bibnamefont
  {Wiese}},\ }\bibfield  {title} {\bibinfo {title} {Simulations of {{Discrete
  Quantum Systems}} in {{Continuous Euclidean Time}}},\ }\href
  {https://doi.org/10.1103/PhysRevLett.77.5130} {\bibfield  {journal} {\bibinfo
   {journal} {Physical Review Letters}\ }\textbf {\bibinfo {volume} {77}},\
  \bibinfo {pages} {5130} (\bibinfo {year} {1996})}\BibitemShut {NoStop}%
\bibitem [{\citenamefont {Banks}\ and\ \citenamefont
  {Zaks}(1982)}]{banks_phase_1982}%
  \BibitemOpen
  \bibfield  {author} {\bibinfo {author} {\bibfnamefont {T.}~\bibnamefont
  {Banks}}\ and\ \bibinfo {author} {\bibfnamefont {A.}~\bibnamefont {Zaks}},\
  }\bibfield  {title} {\bibinfo {title} {On the phase structure of vector-like
  gauge theories with massless fermions},\ }\href
  {https://doi.org/10.1016/0550-3213(82)90035-9} {\bibfield  {journal}
  {\bibinfo  {journal} {Nuclear Physics B}\ }\textbf {\bibinfo {volume}
  {196}},\ \bibinfo {pages} {189} (\bibinfo {year} {1982})}\BibitemShut
  {NoStop}%
\bibitem [{\citenamefont {Sierra}(1997)}]{sierra_application_1997}%
  \BibitemOpen
  \bibfield  {author} {\bibinfo {author} {\bibfnamefont {G.}~\bibnamefont
  {Sierra}},\ }\bibfield  {title} {\bibinfo {title} {On the application of the
  nonlinear sigma model to spin chains and spin ladders},\ }\href
  {https://doi.org/10.1007/BFb0104637} {\bibfield  {journal} {\bibinfo
  {journal} {Lect.Notes Phys.}\ }\textbf {\bibinfo {volume} {478}},\ \bibinfo
  {pages} {137} (\bibinfo {year} {1997})}\BibitemShut {NoStop}%
\end{thebibliography}%
